\documentclass[twocolumn]{aastex63}

\newcommand{\logg}{$\log g$\,}
\newcommand{\teff}{$T_\mathrm{eff}$}
\newcommand{\feh}{[Fe/H]}
\newcommand{\alphafe}{[$\alpha$/Fe]}
\usepackage{mathrsfs}
\usepackage{csvsimple}
\usepackage[normalem]{ulem}

\shorttitle{Coadded abundances in M31 dSphs}
\shortauthors{Wojno et al.}

\begin{document}

\title[Coadded abundances in M31 dSphs]{[Fe/H] and [$\alpha$/Fe] Abundance Measurements in M31 Dwarf Galaxies Using Medium-Resolution, Coadded Spectra from the SPLASH Survey}

\title{Elemental abundances in M31: [Fe/H] and [$\alpha$/Fe]  in M31 Dwarf Galaxies Using Coadded Spectra}

\correspondingauthor{Jennifer Wojno}
\email{jwojno1@jhu.edu}

\author[0000-0002-3233-3032]{Jennifer Wojno}
\affiliation{Department of Physics and Astronomy, Johns Hopkins University, 
3400 N. Charles St,
Baltimore, MD 21218, USA}

\author[0000-0003-0394-8377]{Karoline M. Gilbert}
\affiliation{Department of Physics and Astronomy, Johns Hopkins University,
3400 N. Charles St,
Baltimore, MD 21218, USA}
\affiliation{Space Telescope Science Institute,
3700 San Martin Dr.,
Baltimore, MD 21218, USA}

\author[0000-0001-6196-5162]{Evan N. Kirby}
\affiliation{California Institute of Technology, 1200 E.\ California Blvd., MC 249-17, Pasadena, CA 91125, USA}

\author[0000-0002-9933-9551]{Ivanna Escala}
\affiliation{California Institute of Technology, 1200 E.\ California Blvd., MC 249-17, Pasadena, CA 91125, USA}
\affiliation{Department of Astrophysical Sciences, Princeton University, 4 Ivy Lane, Princeton, NJ~08544}

\author[0000-0002-1691-8217]{Rachael L. Beaton}
\altaffiliation{Hubble Fellow}
\altaffiliation{Carnegie-Princeton Fellow}
\affiliation{Department of Astrophysical Sciences, Princeton University, 4 Ivy Lane, Princeton, NJ~08544}
\affiliation{The Observatories of the Carnegie Institution for Science, 813 Santa Barbara St., Pasadena, CA~91101}

\author[0000-0002-9599-310X]{Erik J. Tollerud}
\affiliation{Space Telescope Science Institute,
3700 San Martin Dr.,
Baltimore, MD 21218, USA}

\author[0000-0003-2025-3147]{Steven R. Majewski}
\affiliation{Department of Astronomy, University of Virginia, Charlottesville, VA 22904-4325, USA}

\author[0000-0001-8867-4234]{Puragra Guhathakurta}
\affiliation{Department of Astronomy \& Astrophysics, University of California, Santa Cruz, 1156 High Street, Santa Cruz, CA 95064, USA}


\begin{abstract}

We present chemical abundances of red giant branch (RGB) stars in the dwarf spheroidal (dSph) satellite system of Andromeda (M31), using spectral synthesis of medium resolution (R $\sim 6000$) spectra obtained with the Keck II telescope and DEIMOS spectrograph via the Spectroscopic and Photometric Landscape of Andromeda's Stellar Halo (SPLASH) survey. We coadd stars according to their similarity in photometric metallicity or effective temperature to obtain a signal-to-noise ratio (S/N) high enough to measure average [Fe/H] and [$\alpha$/Fe] abundances. We validate our method using high S/N spectra of RGB stars in Milky Way globular clusters as well as deep observations for a subset of the M31 dSphs in our sample. For this set of validation coadds, we compare the weighted average abundance of the individual stars with the abundance determined from the coadd. We present individual and coadded measurements of [Fe/H] and [$\alpha$/Fe] for stars in ten M31 dSphs, including the first [$\alpha$/Fe] measurements for And IX, XIV, XV, and XVIII. These fainter, less massive dSphs show declining [$\alpha$/Fe] relative to [Fe/H], implying an extended star formation history. In addition, these dSphs also follow the same mass-metallicity relation found in other Local Group satellites. The conclusions we infer from coadded spectra agree with those from previous measurements in brighter M31 dSphs with individual abundance measurements, as well as conclusions from photometric studies. These abundances greatly increase the number of spectroscopic measurements of the chemical composition of M31's less massive dwarf satellites, which are crucial to understanding their star formation history and interaction with the M31 system.
\end{abstract}

\keywords{galaxy: M31 --- galaxies: dwarf --- galaxies: abundances}

\section{Introduction} \label{sec:intro}

Deriving detailed chemical abundances of stars is a crucial component in modeling the history of galaxies. Chemical abundance patterns across stellar populations provide information for both external \citep[e.g. halo accretion/merger history;][]{Bullock05,cooper10,font11}, as well as internal \citep[e.g. radial migration;][]{gilmore02,Boeche13b,Kordopatis15_rich,Hayden15} evolutionary processes. The ratio of $\alpha$ elements to iron (Fe) is especially illuminating with respect to a galaxy's star formation history. This is because $\alpha$-elements and iron-peak elements are produced on distinctly different timescales, and can therefore act as an indicator for the star formation history in a particular environment \citep{tinsley79,gilmore91,gilmore98}. $\alpha$ elements, such as oxygen and magnesium, are produced predominantly by one source: core-collapse supernovae. The progenitors of these supernovae are massive (M $> 8M_{\sun}$), and have well-known lifespans \citep[$\sim 10^6$ years,][]{arnett96,pagel97,woosley05}. In comparison, iron-peak elements are produced primarily in Type Ia supernovae \citep[][]{tinsley80} on significantly longer timescales \citep[$\gtrsim 10^8$ years,][]{smecker-hane92,nomoto97,Matteucci01,Ishigaki12}. Stars with higher [$\alpha$/Fe] ratios formed from gas enriched by core-collapse supernovae, while stars with lower [$\alpha$/Fe] ratios formed from gas with a higher iron content -- after Type Ia SNe released their ejecta into the ISM\@. In this way, the ratio of [$\alpha$/Fe] and trends in [$\alpha$/Fe]--[Fe/H] space provide clues to disentangle the enrichment timescale of a given population, the environment in which it formed, and its initial mass function. 

Before the era of large-scale, ground-based spectroscopic surveys, such abundance measurements have been limited to stars within the Milky Way (MW), or MW satellites. It is now possible to push beyond the boundaries of our own Galaxy, and determine spectroscopic metallicities for red giant stars in our nearest neighbour, Andromeda (M31) \citep[see][for the most recent such studies]{escala19b,escala18,gilbert19a}. At a distance of 760 kpc \citep{Tully13}, the Andromeda galaxy is the closest grand-design spiral galaxy to the MW, and provides an opportunity to study in detail the formation history of a massive galaxy external to our own. Distinguishing the difference in the  chemical evolution, in particular [$\alpha$/Fe], between these galaxies is crucial in addressing a number of open questions relating to differences in the accretion history of their respective halos \citep[e.g.][]{Ibata94,Ibata01,Carollo07,Ibata13,Helmi17}, their star formation history, the differences their morphology (e.g. bulge-to-disk ratios), and exploring the universality of $[\alpha$/Fe] trends in massive galaxies. 

\citet{vargas14a} produced the first individual metallicity and abundance ([Fe/H] and [$\alpha$/Fe]) measurements of stars in nine M31 satellites. Their sample consisted of red giant branch (RGB) stars in M31 dwarf spheroidal (dSph) satellites, observed with the Deep Imaging Multi-Object Spectrometer \citep[DEIMOS;][]{faber03} instrument on Keck II, using the medium-resolution ($R \sim 6000$) 1200~l~mm$^{-1}$ (1200G) grating, { covering the wavelength region 6300~\AA~$<~\lambda~<~9100$ \AA\@}. To derive individual abundance measurements, they used a spectral synthesis method first described by \citet{kirby08a}, and further developed by \citet{vargas13}, where observed spectra are compared to a grid of synthetic spectra. Across their sample of nine M31 satellites, they found a large range of average \alphafe, with no correlation between \alphafe\ and internal kinematics, stellar density, or proximity to M31. 

To investigate the abundance distribution of the M31 halo, \citet{vargas14b} identified a sample of four stars in M31 dSph fields that were highly likely to belong to the M31 halo. They compared the abundance pattern of their stars to those in the Milky Way halo and found rough agreement between the M31 outer halo and the MW halo. More recently, \citet[][hereafter E19]{escala18} expanded on this sample, producing individual abundance measurements for eleven stars in the M31 halo, located at 23 projected kpc from the center of M31. This sample was obtained using DEIMOS's 600~l~mm$^{-1}$ (600ZD) grating, which offers the advantage of larger wavelength coverage (4500 \AA $< \lambda < 9000$ \AA), as well as increased signal-to-noise (S/N) per pixel compared to the 1200G grating for a given exposure time. 

Using the same abundance pipeline, \citet{escala19b} measured \feh\ and \alphafe\ in four additional M31 fields: the smooth outer halo, inner halo, the Giant Stellar Stream (GSS) and its kinematically cold substructure, and the outer disk. Their final sample, consisting of 70 RGB stars across all four fields, was used to illustrate differences in the accretion histories between the inner and outer halo components, as well as the effect of a major merger on the star formation rate in the disk of M31. \citet{gilbert19a} focused on the GSS, and presented \feh\ measurements for 61 stars, with \alphafe\ measurements for 21 of these. They compared their abundance measurements to M31 dSphs, investigating the progenitor mass of the GSS. Finally, \citet{kirby20} presented \feh\ and \alphafe\ for 256 RGB stars in five M31 dSphs: And I, And III, And V, And VII, and And X. They compare their abundance measurements with MW dwarf satellites, as well as the smooth halo of M31 and the GSS.

As made clear from previous papers in this series \citep{escala19b,escala18,gilbert19a,kirby20}, significant progress is being made in determining individual abundances for stars in M31 and its satellite galaxies. However, sample sizes are not yet large enough to provide maximally informative constraints on the star formation history of M31. Obtaining high enough S/N to measure \feh\ and \alphafe\ for individual stars requires a significant time investment. At the distance of M31, using a 10-m class telescope such as Keck, obtaining high enough S/N for RGB spectra to determine chemical abundances requires on the order of a few hours exposure time. As a result, a large majority of the spectra gathered for individual stars is below the threshold of signal to noise required to measure chemical abundances for individual stars. 

However, \citet{yang13} explored the possibility of coadding similar spectra to obtain average chemical abundances (\feh, \alphafe). They tested a coaddition procedure using RGB stars in MW globular clusters (GCs) and dwarf spheroidal galaxies (dSphs), using 1200 l mm$^{-1}$ DEIMOS spectra. For stars with high enough S/N to obtain individual measurements, they compared the results of their pipeline to other high quality measurements in the literature, finding good agreement. They also compared the inverse-weighted average \feh\ and \alphafe\ to the results from coadded stars. They proposed that this method can be extended to measure average \feh\ and \alphafe\ from coadded spectra in M31, primarily using spectra that do not have the S/N needed to measure chemical abundances.

In this study, we extend the measurement of \feh\ and \alphafe\ abundances through spectral synthesis to coadded spectra in M31. We update the pipeline described in \citet{yang13} by implementing it in Python, and take into account additional fit uncertainties that result from adopting photometric stellar parameters in the spectral synthesis. The majority of large scale spectroscopic surveys in M31 so far have been focused on obtaining radial velocity measurements to study the kinematics of substructure in the halo, the internal velocity dispersions of M31 satellites, and the dynamics of the M31 disk \citep[][]{tollerud12, tollerud2013, collins2011, collins2013,gilbert18}. The Spectroscopic and Photometric Landscape of Andromeda's Stellar Halo \citep[SPLASH,][]{guhathakurta05, guhathakurta06,gilbert06} survey has collected tens of thousands of RGB spectra throughout the M31 halo, disk, and its satellites \citep[e.g.,][]{kalirai10, dorman12, dorman15, gilbert12, gilbert14, tollerud12}. Many of these spectra do not have the S/N needed to determine metallicity and abundances for individual stars, and are therefore the perfect candidates for coaddition.

This paper is organized as follows. In Section \ref{sec:observations}, we describe the observation and instrument configurations used to gather the data used in our analysis. Section \ref{sec:individual_abundances} gives a brief outline of the chemical abundance pipeline used to determine individual \feh\ and \alphafe\ measurements for RGB stars. Section \ref{sec:coadded_abundances} describes the selection criteria for our sample of stars to be coadded, as well as changes made to this pipeline to handle coadded spectra. In this section we also describe the coaddition process. In Section \ref{sec:uncertainties}, we show results for validation tests used to verify that our coaddition method produces similar results to the weighted average abundance for a group of spectra, and how we determine the overall uncertainties on our coadded abundance measurements. In Section~\ref{sec:coadd}, we present results from our coaddition pipeline using spectra for which individual abundance measurements had large uncertainties, and compare them to our sample of validation coadds. We present the 2D \alphafe-\feh\ distribution in Section~\ref{sec:alphafe_feh}, and discuss the implications of our findings in the larger context of the star formation history of M31 and its satellites. Finally, Section \ref{sec:conclusions} summarizes our results, and describes potential future applications of our coaddition process.

\section{Observations and Primary Data Reduction}
\label{sec:observations}
We use spectra of red giant stars in six Milky Way globular clusters \citep[average S/N across the whole sample of 158.13 per angstrom, ][]{kirby08a,kirby10,kirby16}, and ten M31 satellite galaxies observed as part of the SPLASH survey \citep{tollerud12,gilbert12, gilbert18}, obtained using the DEIMOS instrument \citep{faber03} on Keck II.

{\subsection{Target Selection}
The method for selecting candidate RGB stars in M31 is described in detail in \citet{gilbert06}, \citet{guhathakurta06}, \citet{kalirai06}, and \citet{tollerud12}. 
Shallow ($\sim 1-2$ hr) observations were obtained first (see Table~\ref{tab:sample_dsphs}), {where stars  likely to be RGB stars at the distance of M31 based on their CMD position were given the highest preference during the slitmask design process}.  The target selection for the deep ($\sim 6$ hr exposure) masks was further refined using the shallow spectra to preferentially select stars highly likely to be M31 dSph members. 
Additional details regarding the design of the shallow masks can be found in \citet{kalirai09,kalirai10} and \citet{tollerud12}. 
The imaging used for target selection for the shallow masks targeting And I, II, III, V, VII, IX, and XIV was obtained using the Washington $M$ and $T_2$ filters, as well as the DDO51 intermediate filter \citep{ ostheimer03,kalirai06,tollerud12,beaton14}.
 And X was discovered using photometry from the Sloan Digital Sky Survey (SDSS) \citep{adelman-mccarthy06}, and masks were designed using data from \citet{zucker07}. Target selection for And XV was done using Canada-France-Hawaii Telescope archival imaging \citep{tollerud12}. The mask for And XVIII was designed using $B$ and $V$-band imaging from the Large Binocular Telescope (LBT) \citep{beaton14}.}

{For the slitmasks with deep observations, the photometry used to design the slitmasks was obtained from observations conducted with the Mosaic camera on the Kitt Peak National Observatory (KPNO) 4-meter telescope \citep{beaton14,kirby20}. Imaging was obtained in the Washington $M$ and $T_2$ bands. The photometric transformations relations from \citet{majewski00} were used to transform photometry from the Washington $M$ and $T_2$ magnitudes to Johnson-Cousins $V$ and $I$. A star's position in ($I$, $V$ - $I$) color-magnitude space was then used as part of the criteria to identify M31 RGB stars from MW field stars (see Section~\ref{sec:membership_quality_criteria}). In addition, photometry was obtained using the intermediate-width, surface-gravity sensitive DDO51 filter. Objects were assigned a DDO51 parameter, corresponding to the likelihood of the object being an RGB star, based on its position in ($M$ - DDO51) - ($M$ - $T_2$) color-color space \citep{palma03}. In general, objects with point-like morphology and high values of the DDO51 parameter were given the highest priority when constructing the spectroscopic slit masks for DEIMOS. }

\subsection{Observations}
 For five of these galaxies (And I, III, V, VII, and X), we have deep ($\sim 6$ hour total exposure time, $\langle \mathrm{S/N} \rangle = 15.3$ per angstrom) spectroscopic observations. For all ten M31 dSphs, we have shallow spectroscopic observations ($\sim 1-2$ hour total exposure time, $\langle \mathrm{S/N} \rangle = 7.6$ per angstrom). We note that these shallow masks were originally designed for obtaining kinematic information only, while the deep masks were explicitly designed to get high enough S/N to determine  abundances for individual stars. Details of the observations for each slitmask can be found in Table~\ref{tab:sample_dsphs}.

Spectra were obtained using the 1200G (1200 l mm$^{-1}$) grating, with the OG550 order blocking filter. The observed spectra cover a spectral range of $\sim6300\mbox{\AA} < \lambda < 9100\mbox{\AA}$, with a resolution of $\sim6000$ at the central wavelength of $\sim7800\mbox{\AA}$. The spectral dispersion of this science configuration is $0.33 \mbox{\AA}$ pix$^{-1}$, with a resolution of 1.2 $\mbox{\AA}$ FWHM. 

\begin{deluxetable*}{llllll}
\tablenum{1}
\tablecaption{MW globular cluster and M31 satellite observations obtained with the  1200G grating on Keck II/DEIMOS.\label{tab:sample_dsphs}}
\tablehead{\colhead{Object name} & \colhead{Mask name} & \colhead{Date} & \colhead{Airmass} & \colhead{$t_{\rm exp}$} & \colhead{N$_{\rm members}$}\\
\colhead{} & \colhead{} & \colhead{} & \colhead{} & \colhead{(s)} & \colhead{}}
\startdata
    MW GCs\tablenotemark{a}               &           &             &         &                   &                   \\
    \hline
    NGC 2419             & n2419c    & 2009 Oct 13 & 1.18    & 4800              & 54                \\
    NGC 6656 (M22)       & n6656b    & 2009 Oct 13 & 1.48    & 4020              & 47                \\
    NGC 1904 (M79)       & 1904l2    & 2014 Aug 28 & 2.37    & 4200              & 18                \\
    NGC 6341 (M92)       & n6341b    & 2011 Jun 2  & 1.1     & 1800              & 26                \\
    NGC 6864 (M75)       & 6864aB    & 2011 Aug 5  & 1.34    & 4800              & 39                \\
    NGC 7078 (M15)       & 7078l1B   & 2014 Aug 28 & 1.02    & 3600              & 41                \\
                         &           &             &         &                   &                   \\
    \hline
    M31 dSphs            &           &             &         &                   &                   \\
    \hline
    Deep observations\tablenotemark{b,c}    &           &             &         &                   &                   \\
    And I                & and1a     & 2016 Dec 29 & ...       & 18300             & 26                \\
    And III              & and3a     & 2016 Sep 7  & ...       & 19059             & 26                \\
    And V                & and5a     & 2016 Sep 27 & ...       & 24275             & 49                \\
                         & and5b     & 2016 Dec 29 & ...       & 17981             & 45                \\
    And VII              & and7a     & 2017 Jan 1  & ...       & 16503             & 74                \\
    And X                & and10a    & 2016 Sep 28 & ...       & 17760             & 12                \\
                         & and10b    & 2017 Jan 1  & ...       & 22165             & 10                \\
                         &           &             &         &                   &                   \\
    Shallow observations &           &             &         &                   &                   \\
    And I                 & d1\_1     & 2005 Nov 05 & 1.07    & 4055              & 17                \\
                         & d1\_2     & 2005 Sep 16 & 1.49    & 3600              & 22                \\
    And II                & d2\_1     & 2005 Sep 06 & 1.36    & 3600              & 51                \\
                         & d2\_2     & 2005 Sep 06 & 1.13    & 3600              & 36                \\
    And III               & d3\_1     & 2005 Sep 08 & 1.56    & 3600              & 27                \\
                         & d3\_2     & 2005 Sep 08 & 1.22    & 3600              & 16                \\
                         & d3\_3     & 2009 Aug 23 & 1.05    & 3600              & 10                \\
    And V                 & d5\_1     & 2008 Sep 30 & 1.13    & 3000              & 29                \\
                         & d5\_2     & 2008 Sep 30 & 1.27    & 2250              & 27                \\
                         & d5\_3     & 2008 Oct 01 & 1.27    & 2250              & 21                \\
    And VII               & d7\_1     & 2008 Aug 04 & 1.37    & 3000              & 65                \\
                         & d7\_2     & 2008 Aug 04 & 1.27    & 1800              & 38                \\
                         & d7\_3     & 2012 Sep 15 & 1.19    & 10800             & 64                \\
    And IX                & d9\_1     & 2009 Aug 26 & 1.09    & 2700              & 21                \\
                         & d9\_2     & 2009 Aug 26 & 1.15    & 2400              & 10                \\
    And X                 & d10\_1    & 2005 Sep 05 & 1.41    & 3600              & 10                \\
                         & d10\_2    & 2005 Sep 05 & 1.2     & 3600              & 11                \\
    And XIV               & A170\_1   & 2006 Nov 20 & 1.24    & 3600              & 14                \\
                         & A170\_2   & 2006 Nov 21 & 1.2     & 3600              & 24                \\
                         & d14\_3    & 2009 Aug 22 & 1.02    & 3600              & 10                \\
    And XV                & d15\_1    & 2009 Aug 25 & 2.18    & 3600              & 18                \\
                         & d15\_2    & 2009 Aug 25 & 1.39    & 4800              & 10                \\
    And XVIII             & d18\_1    & 2009 Aug 26 & 1.3     & 10800             & 20   \\       
    \hline
\enddata
\tablenotetext{a}{Observations of MW GCs were not obtained as part of the SPLASH survey. More details can be found in \citet{kirby08a,kirby10,kirby16}.}
\tablenotetext{b}{For M31 dSphs with deep ($\sim 6$ hour total exposure time) observations, we list the observation date of the last set of observations taken.}
\tablenotetext{c}{Where observations were conducted over a long timescale (e.g. months), we do not give a value for the airmass.}
\end{deluxetable*}

\subsection{Data reduction pipeline}
One-dimensional science spectra were extracted from raw 2D DEIMOS spectra using the DEEP2 DEIMOS data reduction pipeline, contained within the \texttt{spec2d} IRAF package, developed by the DEEP2 Galaxy Redshift Survey\footnote{http://deep.ps.uci.edu/spec2d/} \citep{cooper12,newman13}. Details on modifications made for better performance on bright stellar sources (versus faint galaxies for which the pipeline was written) can be found in \citet{simon07}. Line-of-sight velocities for each star were obtained by cross-correlating observed spectra with the stellar templates provided in \citet{simon07}. To correct for the imperfect centering of a star in the slit, we use the relative positions of the Fraunhofer A band and other telluric absorption features \citep{simon07,sohn07}.

\section{Individual abundance measurements}
\label{sec:individual_abundances}
To determine spectroscopic effective temperature, \feh, and \alphafe\, for individual stars, we adopt the synthetic spectral synthesis method outlined in \citet{kirby08a}, and further refined in \citet{escala18} using a Python-based implementation of the method. An in-depth  comparison between the results from the two pipelines (\citet{kirby08a}, written in IDL, and \citet{escala18}, written in Python) is provided in Appendex~\ref{sec:pipeline_comparison}. The observed spectrum is compared to a grid of synthetic spectra, and the \feh\, and \alphafe\, values from the best fit synthetic spectrum are adopted as our measured value. {We apply this method to all individual spectra in our MW GC and M31 dSph samples, regardless of their S/N, or any other quality criteria (see Section~\ref{sec:membership_quality_criteria}).}

\subsection{Photometric atmospheric stellar parameters} 
Photometric effective temperature ($T_{\mathrm{eff,phot}}$), surface gravity (log $g$), and photometric metallicity ([Fe/H]$_{\mathrm{phot}}$) for individual stars are determined by linearly interpolating between isochrones,  assuming a single distance for a given object. Assuming an age of 14 Gyr and [$\alpha$/Fe] $= +0.3$ dex, we find the best-fit Padova \citep{Girardi16}, Victoria-Regina \citep{vandenberg06}, and Yonsei-Yale \citep{demarque04} isochrones, using a star's position in $(I, V-I)$ space. We take the error-weighted mean of all three isochrone sets as our final measured $T_{\mathrm{eff,phot}}$, log $g$, and [Fe/H]$_{\mathrm{phot}}$. {While there are slight variations between the photometric values derived from the different isochrone sets, they do not significantly affect the abundance measurement \citep{kirby08a}. Additionally, we justify a single age and $\alpha$-enhancement by noting that for our sample, $T_{\mathrm{eff,phot}}$ and log $g$ are largely insensitive to variations in age and assumed [$\alpha$/Fe] of the isochrone set.} For individual stars, $T_{\mathrm{eff, phot}}$ is used as a first approximation in measuring spectroscopic \teff, and log $g$ is held constant throughout the spectroscopic abundance fitting procedure.

\subsection{Synthetic spectral grid}
The construction of the grid of synthetic spectra for the 1200G grating is described in detail in \citet{kirby08a,kirby09} and \citet{kirby11}. To summarize, synthetic spectra are generated using modified ATLAS9 model atmospheres \citep{kirby11, kurucz93,sbordone04,sbordone05} without convective overshooting \citep{castelli97}. The line list for atomic transitions were sourced from the Vienna Atomic Line Database \citep[VALD;][]{kupka99}, molecular lines from \citet{kurucz92}, and the hyperfine transition line list from \citet{kurucz93}, where the oscillator strengths were tweaked to match values from observed spectra of the Sun and Arcturus \citep{kirby11}. The parameter ranges for the stellar atmospheric parameters cover old red giant branch stars, subgiants, and lower main sequence stars: $3500 $ K $\leq$ $T_{\mathrm{eff}}$ $\leq$ 8000 K, $0.0 \leq $ log $g \leq $ 5.0, $-5.0 \leq$ [Fe/H]$\leq 0.0$, and $-0.8 \leq $ [$\alpha$/Fe] $\leq +1.2$ (further details of the grid can be found in \citet{kirby08a,kirby09}). Synthetic spectra are then generated using the LTE stellar synthesis code MOOG \citep{sneden73}, with a resolution of 0.02\mbox{\AA}. 

\subsection{Chemical abundance pipeline}
Here we outline the pipeline we use to measure \feh\ and \alphafe\ for individual stars \citep{kirby08a, escala18}, which we modify to measure abundances from coadded spectra as discussed in Section~\ref{sec:coadded_abundances}.

\subsubsection{Telluric correction}
Spectra are corrected for absorption by the Earth's atmosphere using the method described in \citet{kirby08a}, where the observed spectrum is divided by a scaled telluric absorption template, taken from a spectrophotometric standard star {\citep{simon07}}.  For our spectra, we use BD +28 4211, observed on November 7, 2007 using the same configuration as our primary science observations.

\subsubsection{Initial continuum normalization}
\label{sec:initial_cont_normalization}
The telluric-divided observed spectrum is cross-correlated with a template spectrum to determine radial velocity, as in \citet{kirby15}, {using empirical stellar templates that cover a range of effective temperature, surface gravity, and spectral types.} After spectra have been corrected for telluric absorption and shifted to the rest frame, they are continuum normalized. The continuum is fit using a third-order B spline over the continuum regions \citep{kirby08a} with a breakpoint spacing of 100 pixels for the 1200G grating. A sigma clipping is performed ($5 \sigma$ tolerance, {where $\sigma$ is the inverse square root of the inverse variance array}) to remove pixels that deviate significantly from the continuum fit. These pixels are not included in subsequent iterations of the continuum normalization. {In addition to this initial normalization, we also refine the continuum fit as we solve for \feh\ and \alphafe\ (Section~\ref{sec:pipeline_process}). In each step of the process, the {continuum} fit is weighted by the inverse variance of the observed spectrum.}

\subsubsection{Abundance masks}
\label{sec:abundance_masks}
Abundance masks are constructed for measuring both [Fe/H] and [$\alpha$/Fe] during the fitting procedure, to ensure that only regions sensitive to either [Fe/H] or [$\alpha$/Fe] are considered. We use the same mask regions as those defined in \citet{kirby08a}.
We note that the mask for [$\alpha$/Fe] is constructed from regions sensitive to the individual alpha elements Mg, Ca, Si, and Ti. 
In addition to these abundance masks, a general mask is applied to all spectra to exclude pixels at the beginning and end of the spectra, pixels near the detector chip gap, regions of bad sky subtraction, and other detector artifacts \citep{escala18}. 

\subsubsection{Measuring spectroscopic $T_{\mathrm{eff}}$, [Fe/H], [$\alpha$/Fe]}
\label{sec:pipeline_process}
In the first stage of fitting a synthetic spectrum to the observed spectrum, effective temperature ($T_{\mathrm{eff}}$) and iron abundance ([Fe/H]) are allowed to vary simultaneously. An abundance mask is applied to consider only regions sensitive to changes in [Fe/H]. 

The photometric temperature 
is used as a starting point for the fit, with \feh\, initialized at $-2.0$ dex, and \alphafe\, initialized at $0.0$ dex. We note that the choice of initial values does not affect the final measured abundances \citep{escala18}. We assume the spectral resolution ($\Delta \lambda$) is a function of wavelength, determined from measuring the Gaussian widths of {over 100 approximately evenly spaced} sky lines for each slitmask \citep{kirby08a}. This is a departure from the method presented in  \citet{escala18}, where they fit for $\Delta \lambda$ due to a lack of sky lines at bluer wavelengths in the wavelength range they use in their study ($\sim4500-9100$ \AA).
These parameters ($T_{\mathrm{eff}}$, \feh) are allowed to vary iteratively until a best fit is found. 

In the next step of fitting, the best fit values of \teff\ and \feh\ determined in the previous step are held constant, and \alphafe\ is allowed to vary. A mask is applied to consider only regions sensitive to changes to \alphafe, in an analogous way to the mask for \feh. 

Using the best fit values for $T_{\mathrm{eff}}$, [Fe/H], and [$\alpha$/Fe], the continuum level is then refined by dividing the continuum-normalized observed spectrum by the best fit interpolated synthetic spectrum, and fitting a third-order B-spline to this ``flat-noise" spectrum with 3$\sigma$ clipping. The spline fit is then subtracted from the continuum-normalized observed spectrum. The effective temperature fitting, abundance fitting, and continuum refinement process are repeated iteratively until the best fit values vary by less than 1 K, 0.001 dex, and 0.001 dex for $T_{\mathrm{eff}}$, [Fe/H], and [$\alpha$/Fe], respectively. A flag is included in the final output where this continuum refinement step does not converge after 50 iterations.

The fit uncertainty is calculated from the reduced $\chi^2$ and the covariance matrices of the fit between the best fit synthetic spectrum and the observed spectrum. For the systematic uncertainty, we adopt the same uncertainty floors as \citet{gilbert19a}: 0.101 dex for \feh\ and 0.084 dex for \alphafe. {This systematic uncertainty is determined by comparing abundance measurements from high resolution spectra with those from the medium-resolution DEIMOS spectra for a set of validation stars, and requiring that these values agree within $1\sigma$ \citep{kirby09}.}
The total uncertainty is then determined from adding in quadrature the fit and systematic uncertainties.

For our validation sample, we require that they have ``good'' spectra, i.e., no problems with sky subtraction or artifacts in the spectra that would be apparent upon visual inspection. Additionally, we examine the $\chi^2$ contours of the \teff, \feh, and \alphafe\ fits for each star, and remove any stars that have contours that are not smooth and continuous across the parameter range, and therefore less likely to provide reliable spectroscopic parameters. Finally, for our validation sample (Section~\ref{sec:uncertainties}), we require that stars have uncertainties of less than 0.4 dex in both \feh\ and \alphafe, their spectra converged during the continuum refinement step in the synthetic spectral pipeline, and their measured abundances do not hit the limits of the synthetic spectral grid. 

\section{Coadding Spectra}
\label{sec:coadded_abundances}
After we measure spectroscopic \teff, \feh, and \alphafe\  for all stars in our MW GC and M31 dSph samples using the pipeline described above, we apply a number of membership cuts to ensure that we only select stars that are highly likely to belong to the object in question. In addition, we apply a set of quality criteria to select a set of stars that will be used for validating our pipeline (Section~\ref{sec:coadd_vs_weighted_average}). We refer to the coadds using these stars as ``validation coadds''. Any stars which pass the membership criteria but do not pass the quality criteria we consider our science sample, and we refer to coadds constructed from this sample of stars as our ``science coadds''.

\subsection{Membership and quality criteria}
 \label{sec:membership_quality_criteria}
To ensure membership for stars in our globular cluster sample ($\langle \mathrm{S/N} \rangle = 158.13$ per angstrom), we use the $(I, V-I)$ CMD to select only those stars that fall on the RGB or AGB. We also require that stars satisfy \logg $ < 3.6$ to ensure that they are giants.

\begin{figure*}
 	\includegraphics[width=1.01\textwidth]{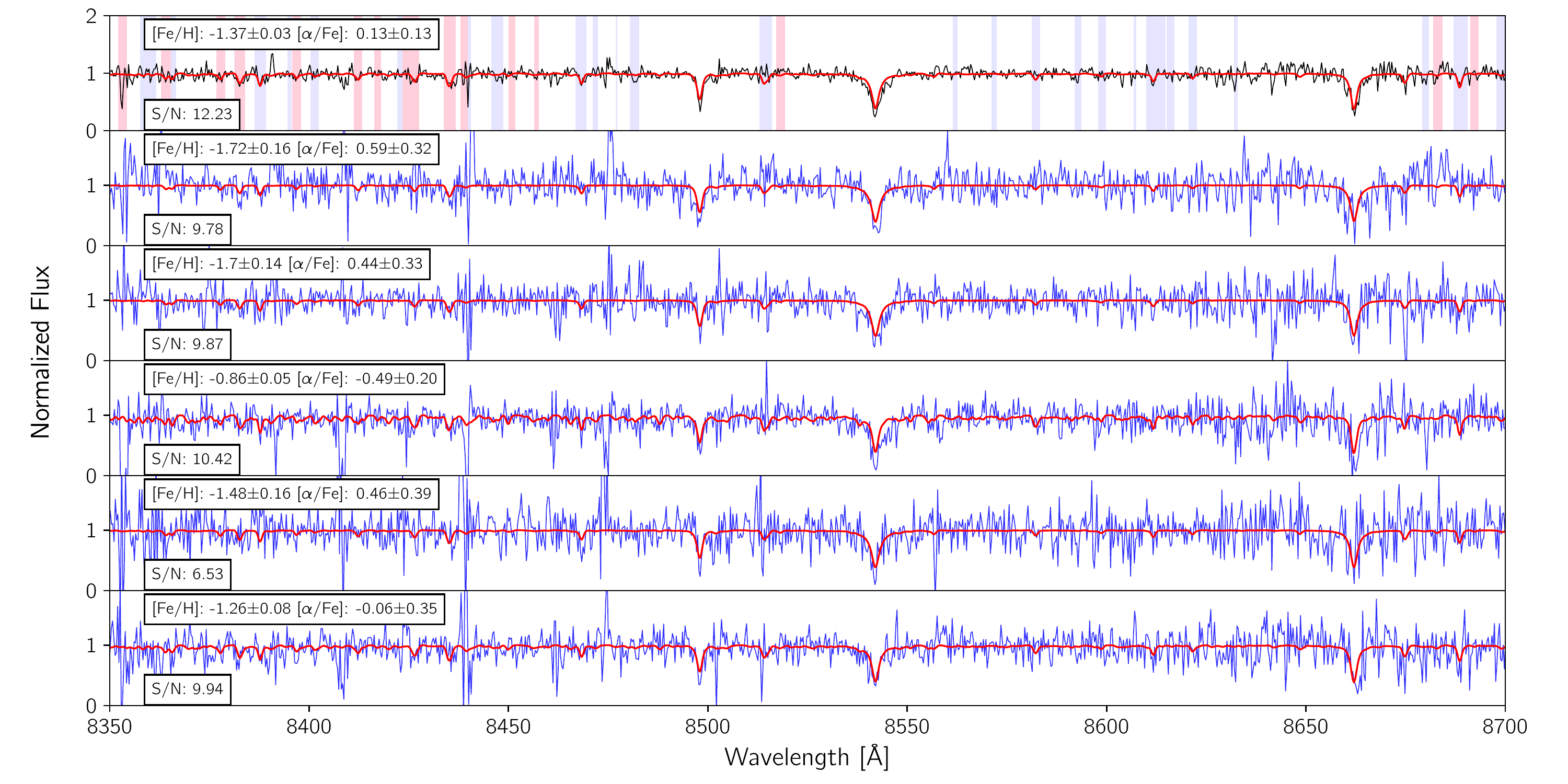}
    \caption{A coadded spectrum (top panel) using spectra from d7\_1, a dSph mask with shallow spectroscopic observations. Signal-to-noise, derived spectroscopic \feh\ and \alphafe\ are given in each panel, with the best fit synthetic spectrum overplotted in red. Regions sensitive to a given abundance and included in the fitting process are indicated in light blue and pink for \feh\ and \alphafe, respectively. Spectra of the five RGB stars used in constructing the coadded spectrum are shown in the bottom five panels in blue. The coadded spectrum illustrates the improvement in S/N (and therefore our ability to measure abundances) over the individual spectra.}
    \label{fig:coadd_example}
\end{figure*}

{For our sample of stars in M31 dSphs, we adopt the criteria used previously for analysis of these samples.  We describe below the differences between the membership selection used for masks with deep observations, and those with shallow observations.}
For our M31 dSphs with deep ($\sim 6$ hr, $\langle \mathrm{S/N} \rangle = 15.3$ per angstrom) observations, {we identify dSph members using a two-step process.  First,} we use the M31/MW membership classification scheme described by \citet{gilbert06} {to identify stars that are likely to be RGB stars at approximately M31's distance, rather than foreground MW dwarf stars}. This membership classification is based on a number of criteria including radial velocity, a star's position in ($M-{\rm DDO51}$) versus ($M-T_2$) color-color space, its position in $(I, V-I)$ color-magnitude space, and the equivalent width of the Na I doublet at 8190 \AA\ ($\lambda \lambda$ 8183,8195). 
We note that the membership classification for our dSphs assumes M31's distance modulus. {Therefore, we consider stars to be a dSph member if they are more probable to be an M31 RGB star than a MW foreground star based on the above diagnostics. We make an exception for stars that are slightly too blue for the most metal-poor isochrone at the distance of M31, and would otherwise be classified as MW foreground stars.} We include these stars in our membership selection because they are likely to belong to the M31 satellite if they pass additional velocity cuts \citep{gilbert06}. {Second,} we require that all stars that satisfy the {M31 RGB} criteria also explicitly fall within the expected radial velocity limits for a given M31 dSph, listed in Table~\ref{tab:velocity_cuts}. These velocity limits correspond to approximately the difference between the mean systemic velocity and three times the velocity dispersion for a given dSph \citep[see Table 3 of ][]{kirby20}.

For M31 dSph stars with shallow ($\sim1$-2 hour, $\langle \mathrm{S/N} \rangle = 7.6$) observations, we adopt the {M31 dSph} membership criteria described by \citet{tollerud12}. This membership determination {simultaneously evaluates the probability the star is an RGB star at the distance of the M31 system, as well as a member of the dSph.  It relies on three {probabilistic} criteria: the distance between a given star and the dSph to which it lies near, a star's position in ($T_2$, $M-T_2$) color-magnitude space, and the equivalent width of the Na I doublet. In addition, a ``wide selection window around the dSph’s systemic velocity" \citep{tollerud12} is used to remove obvious outliers from their dSph membership. These four criteria are then used
to calculate the  probability of a star being a member of a given dSph \citep[see Eq. 1 of][]{tollerud12}, where stars having a membership probability greater than 0.1 are selected as dSph members.} We include the systematic velocities of these dSphs with shallow observations in Table~\ref{tab:velocity_cuts} for completeness. 

\begin{deluxetable}{llllc}
\tablenum{2}
\tablecaption{Kinematic properties of M31 dSphs.\label{tab:velocity_cuts}}
\tablehead{\colhead{Object name} & \colhead{$\langle v_{\mathrm{helio}}\rangle$} & \colhead{$\sigma_{v}$}  & \colhead{Velocity Limits} & \colhead{Reference\tablenotemark{a}} \\
\colhead{} & \colhead{(km s$^{-1}$)} & \colhead{(km s$^{-1}$)} & \colhead{(km s$^{-1}$)} & \colhead{}}
\startdata
    And I & $-378.0^{+2.4}_{-2.3}$ & $9.4^{+1.7}_{-1.5}$ & $[-390,-340]$ & 1\\
    And II & $-192.4\pm 0.5$ & & & 2\\
    And III & $-348.0\pm^{+2.7}_{-2.6}$ & $11.0^{+1.9}_{-1.6}$ & $[-370,-320]$  & 1\\
    And V & $-397.7\pm1.5$ & $11.2^{+1.1}_{-1.0}$ & $[-440,-360]$ & 1 \\
    And VII & $-311.2\pm1.7$ & $13.2^{+1.2}_{-1.1}$& $[-340,-260]$ & 1\\
    And IX & $-209.4\pm2.5$ & & & 3 \\
    And X & $-166.9\pm1.6$ & $5.5^{+1.4}_{-1.2}$ & $[-180,-140]$ & 1\\
    And XIV & $-480.6\pm1.2$ & & &  3 \\
    And XV & $-323.0\pm1.4$ & & &  3 \\
    And XVIII & $-332.1\pm2.7$ & & &  3 \\
    \hline
\enddata
\tablenotetext{a}{References. 1: \citet{kirby20} (deep observations), 2: \citet{ho12} (shallow observations), 3: \citet{tollerud12} (shallow observations)}
\end{deluxetable}

\subsection{Selecting groups for coaddition}
We adopt two binning strategies:\@ sorting stars by either photometric \teff\ or photometric metallicity ([Fe/H]$_{\mathrm{phot}}$), and then binning them such that each coadd has at least five stars. We find, as in \citet{yang13}, that the final measured abundances are consistent when stars are grouped by either photometric temperature or metallicity, so for the purposes of our analysis we simply use the photometric metallicity. The choice to bin by  photometric \feh\ is motivated by the reasoning given in \citet{yang13}, where they state that the synthetic spectra can account for a range of \teff\ within a bin, but not \feh\ or \alphafe. For completeness, we include a brief summary of the following analysis done with stars sorted by photometric \teff\ in Appendix~\ref{sec:bin_teff}. 

A group of spectra to be coadded are then processed using the same method described above (Section~\ref{sec:individual_abundances}), where they are corrected for telluric absorption and shifted to the rest frame, and an initial continuum normalization is performed. Each spectrum in the group is then rebinned to the same spectral range ($6300-9000$~\AA), with a wavelength spacing of 0.32~\AA\,for the 1200 l mm$^{-1}$ grating, to ensure that all spectra are on the same pixel array. This rebinning is applied to both the flux and inverse variance arrays for each spectrum. We then perform a two-step sigma clipping for each pixel in the group of spectra to be added. If a pixel for any spectrum in the group deviates more than 10$\sigma$ or 3$\sigma$ (where $\sigma$ is the standard deviation of the flux for a given pixel) from the median flux for the first and second step of the sigma clipping, respectively, the pixel is masked when adding that spectrum to the coadded spectrum. 

\begin{figure*}
 	\includegraphics[width=0.90\textwidth]{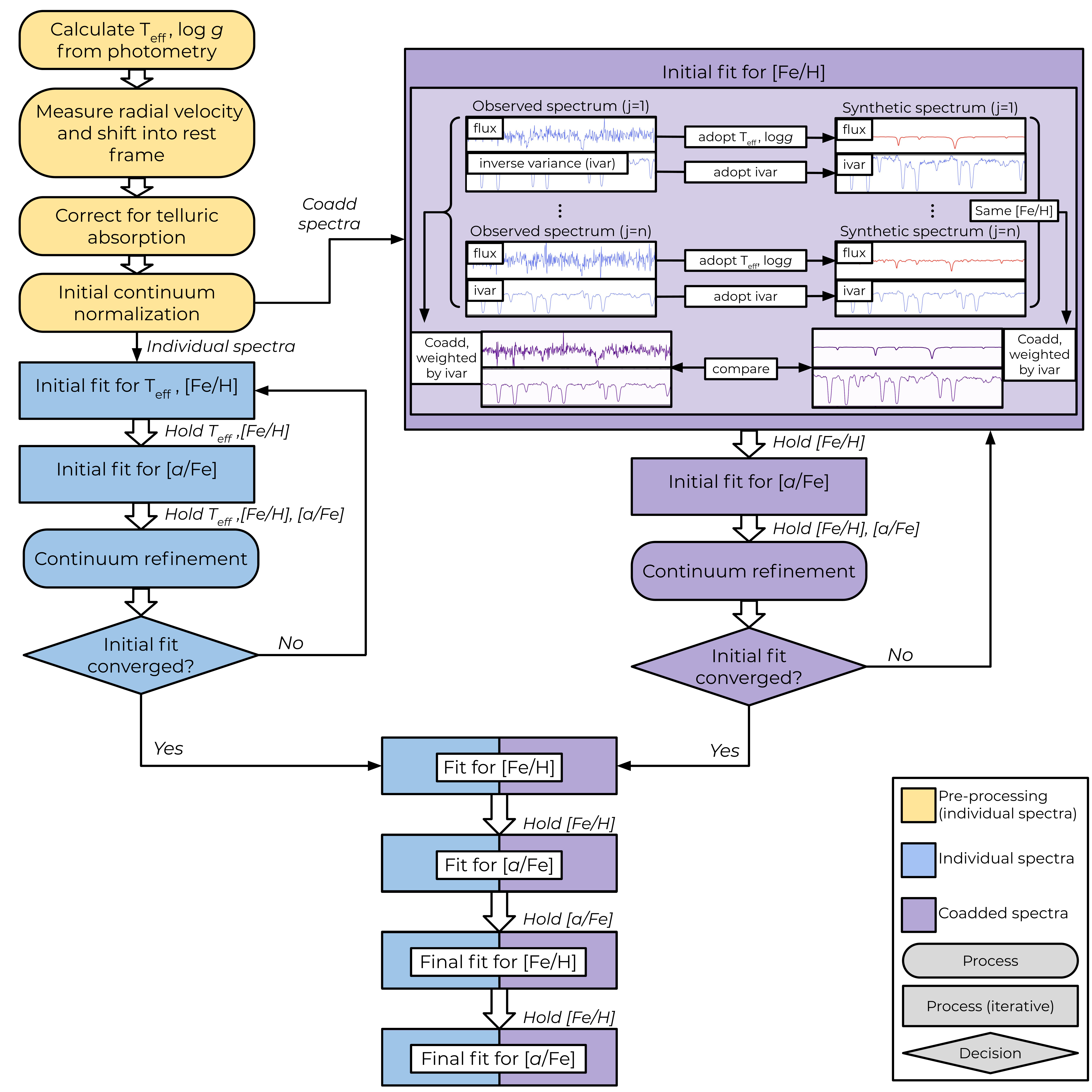}
    \caption{A step-by-step illustration of our process to determine \feh\ and \alphafe\ from individual (blue) and coadded (purple) spectra. Pre-processing steps are shown in yellow. Rounded rectangles indicate single-step processes, while rectangles indicate processes that involve iterating through a loop (i.e., fitting procedures). Steps indicated with a diamond represent an \textit{if} statement. The inset on the top right shows how synthetic spectra are constructed by adopting the photometric effective temperature, surface gravity, and inverse variance array of the corresponding observed spectra.}
    \label{fig:pipeline_process}
\end{figure*}

\subsection{Coaddition}
Fluxes are coadded together weighted by their inverse variance $(1/\sigma^2)$: 
\begin{equation}
\label{eq:coadd}
    \bar{x}_{\mathrm{pixel}, i} = \frac{\sum^{n}_{j=1} (x_{\mathrm{pixel},ij}/\sigma{^2}_{\mathrm{pixel},ij})}{\sum^{n}_{j=1} 1/\sigma{^2}_{\mathrm{pixel},ij}}
\end{equation}
where $x_{\mathrm{pixel},ij}$ represents the flux of the ith pixel of the jth spectrum in a group,  $\sigma{^2}_{\mathrm{pixel},ij}$ is the corresponding  variance of $x_{\mathrm{pixel},ij}$, and $n$ is the total number of spectra to be coadded. The inverse variance  weighted flux of the ith pixel of the coadded spectra is then given by $\bar{x}_{\mathrm{pixel},i}$. The inverse variance of the coadded spectra is: 
\begin{equation}
    \label{eq:inv_var}
    \bar{\sigma}_{\mathrm{pixel},i} = \left( \sum^{n}_{j=1} \frac{1}{\sigma{^2}_{\mathrm{pixel},ij}} \right)^{-1/2}
\end{equation}

Figure~\ref{fig:coadd_example} shows a small portion of the full spectral range for an example coadd, where the final coadded spectrum and the corresponding best-fit synthetic spectrum (red) are shown in the top panel, and the individual constituent spectra are shown in the lower panels.

\subsection{Measuring abundances from coadded spectra}
\label{sec:measuring_abundances_coadd}
To ensure an accurate comparison between our observed coadded spectra and the synthetic spectral grid, we coadd synthetic spectra using the same procedure as the observed spectra (Eq.~\ref{eq:coadd}). During each abundance determination step, {from the synthetic spectral grid, we select a group of synthetic spectra with temperatures and surface gravities that correspond to the \teff\ and \logg\ (determined from photometry) for each observed spectrum}. The synthetic spectra are rebinned to the same wavelength range and pixel scale as the observed spectra, and are smoothed with a Gaussian filter to replicate the spectral resolution of the observed spectra, where the method to determine the spectral resolution is described in Section~\ref{sec:pipeline_process}. The synthetic spectra are then weighted by the same inverse variance array (Eq.~\ref{eq:inv_var}) as their corresponding observed counterparts to produce a coadded synthetic spectrum for each iteration of the fitting procedure.

The process for measuring abundances from coadded spectra varies slightly from the process outlined in Section~\ref{sec:individual_abundances}. When finding the best fit between our coadded observational spectrum and coadded synthetic spectra, we use photometric \teff\ and \logg\ estimates from each observed star for their counterpart in the coadded synthetic spectrum in each iteration of the fit. We first fit only [Fe/H], keeping [$\alpha$/Fe] fixed, and then fit [$\alpha$/Fe], keeping [Fe/H] fixed, using the same masks described in Section~\ref{sec:abundance_masks} for these abundance measurements. We refine the continuum of the coadded spectrum as described in Section~\ref{sec:pipeline_process}. We iterate over these  three steps (fitting \feh, \alphafe, and continuum refinement) until the measured values for [Fe/H]\, [$\alpha$/Fe] change by less than 0.001 dex on consecutive passes.

We then take the coadded spectrum with its refined continuum and fit for \feh, taking the \alphafe\ value from the last iteration of the continuum refinement. We repeat this process for \alphafe, keeping \feh\ constant from the previous step, and adopt this value as our final measured \alphafe. Finally, we refit for \feh\ using the final  measured \alphafe, and take the value of \feh from this step as our final measured \feh. After this round of fitting \feh\ and \alphafe, we check the goodness of the fit. For our coadded abundance measurements, we apply the same quality criteria as in Section~\ref{sec:individual_abundances}. In addition, we examine the $\chi^2$ contours for \feh\ and \alphafe\ to ensure the minimum is well-defined.

\section{Pipeline validation and estimating uncertainties for coadded spectra}
\label{sec:uncertainties}

\begin{figure}
	\includegraphics[width=0.9\columnwidth]{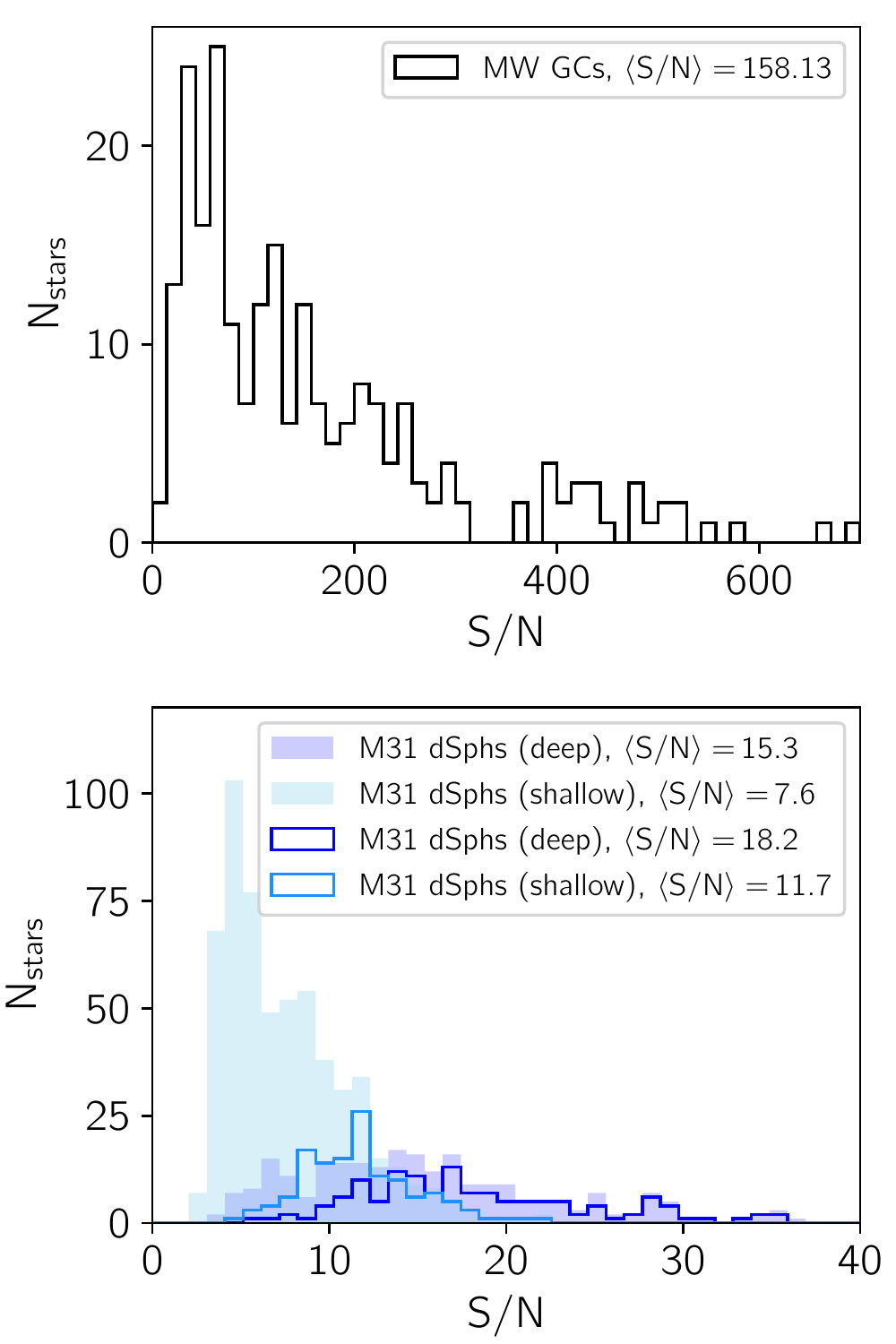}
    \caption{Histograms showing the distribution of S/N (per angstrom) of individual stars in our Milky Way globular cluster sample (top) and our validation M31 dSph stars (bottom). The solid-filled histograms represent the entire sample of member stars, with the lines showing the sample used for validating our coaddition pipeline (Section~\ref{sec:coadd_vs_weighted_average}).}
    \label{fig:sample_snr_hist}
\end{figure}

Validation was done using the six MW globular clusters (NGC 2419, NGC 6656, NGC 1904, NGC 6341, NGC 6864, NGC 7078), four M31 dSphs that have both deep observations and shallow observations (And I, And III, And V, And VII), one dSph that has validation stars from deep observations only (And X), and one M31 satellite with shallow observations only (And II). The number of individual stars and validation coadds for each slitmask for each of these three samples can be found in Table~\ref{tab:sample_validation}. Here, we define our ``validation coadds'' as coadds composed of stars that individually have both \feh\ and \alphafe\ uncertainties less than 0.4 dex, and have a $\chi^2$ minimum that is not at the edge of the synthetic spectral grid ($-5.0$ or 0.0 for \feh, $-0.8$ or 1.2 for \alphafe). {In total, our validation sample consists of 37 validation coadds from GC masks, 24 from M31 dSph masks with deep observations, and 9 from M31 dSph masks with shallow observations.} Figure~\ref{fig:sample_snr_hist} shows the distribution in S/N for each of our three samples, illustrating the S/N regime probed by each. The MW GC sample contains the highest quality spectra with respect to S/N, and provides a first check for the efficacy of our pipeline. The M31 dSph samples allow us to test the coaddition pipeline at lower S/N regimes,
where different sources of noise may hinder our ability to determine chemical abundances. In particular, the M31 dSph sample from shallow observations provides a probe of the lowest S/N at which we can reliably measure chemical abundances from individual spectra {(S/N $\gtrsim 8$ per Angstrom, although we do not apply a strict S/N cut for individual spectra to any of our samples)}.

\begin{table}
    \renewcommand\thetable{3}
	\centering
	\caption{Validation Sample}
	\label{tab:sample_validation}
	\begin{tabular}{llrrr}
	\hline
	\hline
	Object name & Slitmask & N$_{\mathrm{stars}}$ & N$_{\mathrm{coadds}}$\\
    \hline
    MW GCs & & & & \\
    \hline
	NGC 2419 & n2419c  & 49 & 9\\
	NGC 6656 (M22) & n6656b & 35 & 7\\
	NGC 1904 (M79) & 1904l1 & 18 & 3 \\
	NGC 6341 & n6341 & 25 &  5\\
	NGC 6864 & n6864 & 36 &  7\\
	NGC 7078 (M15) & 7078l2 & 33 & 6\\
	\hline
    M31 dSphs & & & & \\
    \hline
    Deep observations & & & \\
	AndI & and1a & 16 & 3\\
	AndIII & and3a & 17 & 3 \\
	AndV & and5a & 28 & 4 \\
	& and5b & 23 & 4 \\
	AndVII & and7a & 58 & 9 \\
	AndX & and10b & 3 & 1 \\
    & & & & \\
    Shallow observations & & & & \\
    AndI & d1\_1 & 7 & 1 \\
     & d1\_2 & 11 & 1 \\
    AndII & d2\_1 & 15 & 2 \\
    AndIII & d3\_1 & 7 & 1 \\
    AndV & d5\_1 & 9 & 1 \\
    AndVII & d7\_1 & 16 & 3 \\
    
	\hline
	\end{tabular}
\end{table}

\subsection{Uncertainties on coadded spectra}
For our coadded spectra, we consider two sources of uncertainties on \feh\ and \alphafe. First, we estimate the fit (or statistical) uncertainty from the diagonals of the covariance matrix and the reduced chi-square value ($\chi^2$) of the fit. For either \feh\ or \alphafe, we take into account the same spectral mask (Section~\ref{sec:abundance_masks}) used in the fit, in order to only consider regions of the spectrum sensitive to either \feh\ or \alphafe. We also consider the effect of uncertainties on the photometric \teff\ and \logg values as part of the fit uncertainty. The process used to estimate this portion of the fit uncertainty is described in Section~\ref{sec:mc_uncertainties}. Second, we consider the systematic uncertainties, which encompass all other sources of unknown uncertainty in our fits (Section~\ref{sec:systematic_uncertainties}).

\subsubsection{Estimating fit uncertainties due to uncertainties from photometric stellar parameters}
\label{sec:mc_uncertainties}
In measuring \feh\ and \alphafe\, for each star we adopt values for \teff\ and \logg\ derived from photometry. These values are propagated through our pipeline to the comparison with synthetic spectra (Section~\ref{sec:measuring_abundances_coadd}).
 As we do not fit for \teff\ or \logg, we  estimate the component of the fit uncertainty due to uncertainties in these  photometric stellar parameters. We assume the photometric \teff\ and \logg\ uncertainties are Gaussian, and draw 50 random samples from the covariant \teff--\logg\ distributions for each star in a given coadd. We then remeasure \feh\ and \alphafe\ for each of these samples for each coadd. We then take the standard deviation of the abundance measurements from all samples for a given coadd as the typical uncertainty due to the photometric \teff\ and \logg ($\sigma_{\mathrm{MC}}$), and add this uncertainty in quadrature to the fit uncertainty ($\sigma_{\mathrm{fit}}$) to get our total fit uncertainty for both \feh\ and \alphafe:
 
 \begin{equation}
     \sigma_{\mathrm{fit, total}} = \sqrt{\sigma_{\mathrm{fit}}^2 + \sigma_{\mathrm{MC}}^2}
 \end{equation}
 
 The magnitude of this uncertainty is small compared to the fit uncertainty, with median values of approximately 0.02 dex for \feh\ and 0.01 dex for \alphafe.

\subsubsection{Quantifying systematic uncertainties}
\label{sec:systematic_uncertainties}

\begin{figure}
\centering
	\includegraphics[width=0.9\columnwidth]{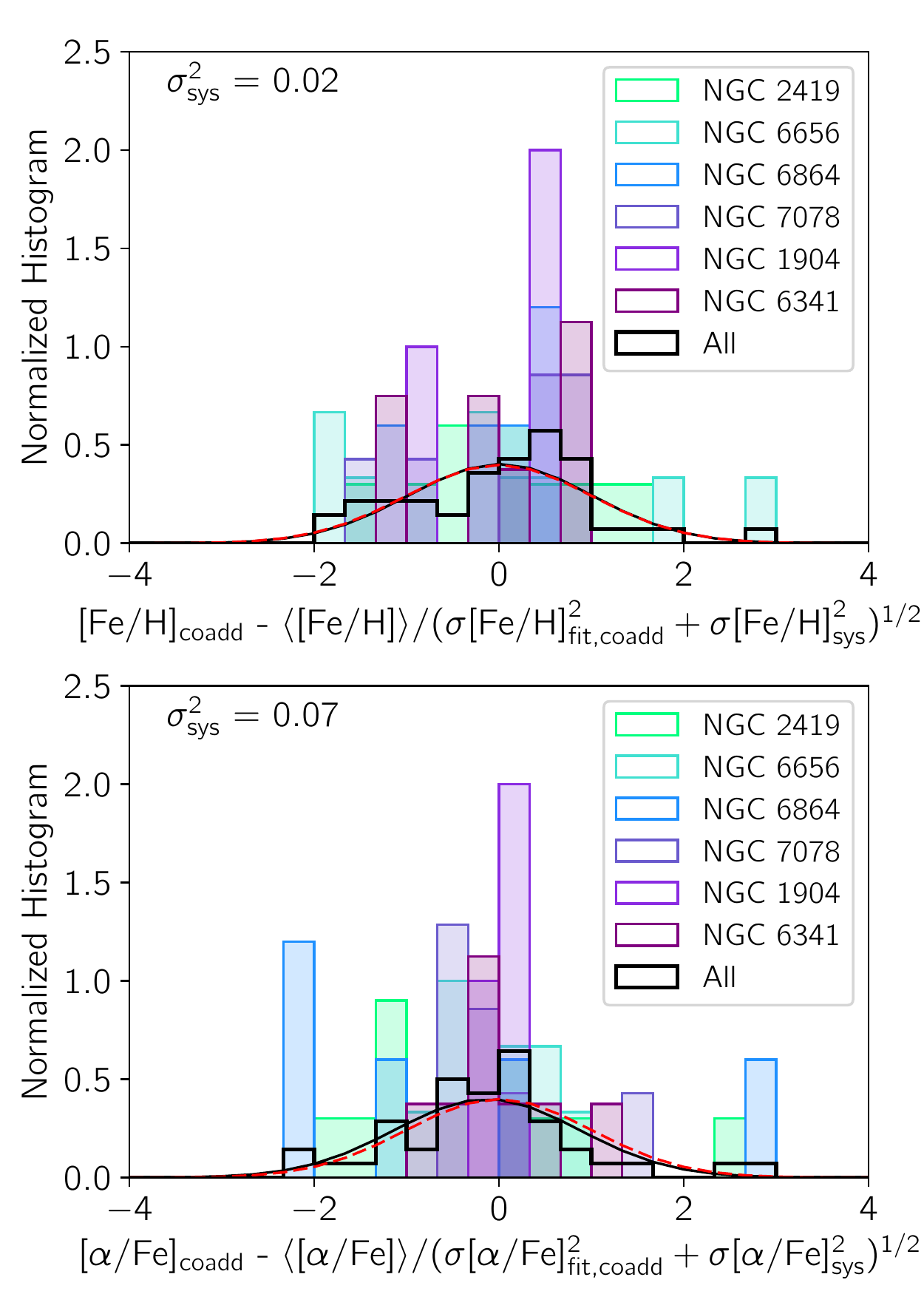}
    \caption{Histogram of the differences between coadded abundances and weighted averages, weighted by their corresponding uncertainties, for [Fe/H] (top) and [$\alpha$/Fe] (bottom). The solid black histogram represents the normalized sum of all validation coadds for all GCs in our sample. The solid black curve indicates the best-fit Gaussian for the sample of all validation coadds from all GCs, and the dashed red curve indicates a normalized unit Gaussian with $\sigma = 1$.}
    \label{fig:uncertainties_hist}
\end{figure}

We estimate the systematic uncertainties on our coadded abundance measurements by estimating the uncertainty term needed to enforce that the measurement from the coadd agrees with the weighted average within one standard deviation. We compute this using all validation coadds from all GCs:

\begin{equation}
\label{eq:sys_uncertainty}
	\frac{1}{\sum_i N_i - 1} \sum_{i=1}^{N_{\rm GC}}  \sum_{j=1}^{N_{\rm i}} \frac{(X_{\mathrm{coadd,i,j}}- \langle X_{i}\rangle)^2}{\sigma_{\mathrm{fit,coadd,i,j}}^2(X) +\sigma_{\mathrm{sys}}^2(X)}= 1
\end{equation}

\noindent where $i$ indicates a given GC, $j$ indicates a given coadd for that GC, $N_{GC}$ is the number of GCs in our sample (6), and $N_i$ is the number of coadds in the $i$th GC. 
The coadded abundance value in a given GC for X = \feh\ or \alphafe\ is $X_{\mathrm{coadd,i,j}}$, $\sigma_{\mathrm{coadd,i,j}}(X)^2$ is the fit uncertainty for that measurement, and $\langle X_{\mathrm{i}}\rangle$ is the weighted average abundance from the coadds for a given GC\@. We then numerically solve for $\sigma^2_{\mathrm{sys}}(X)$, for both \feh\ and \alphafe. With our sample of GCs, we measure a systematic uncertainty of 0.02 dex for \feh\ and 0.07 for \alphafe. Figure \ref{fig:uncertainties_hist} shows the uncertainty-weighted \feh\ and \alphafe\ distributions for the validation coadds in each GC (colored histograms), and for all GCs (black histogram). A Gaussian with the standard deviation equal to the derived systematic uncertainty is overplotted. {The width of this Gaussian corresponds to how well the expected uncertainties capture the total uncertainties in our sample. A best-fit Gaussian that is narrower than a unit Gaussian would indicate that we overestimate our uncertainties, while a best-fit Gaussian that is wider than a unit Gaussian would indicate that we underestimate our uncertainties.}


We find the systematic uncertainty for our GC sample to be of the order of their fit uncertainties. Therefore, we adopt the same systematic uncertainty floors as in \citet{gilbert19a} for spectra measured with the 1200G grating: 0.101 dex for \feh, and 0.084 dex for \alphafe. The total uncertainty on our coadded abundance measurements ($\sigma_{\mathrm{total,coadd}}(X)$) is then:
\begin{equation}
    \sigma_{\mathrm{total,coadd}}(X) = \sqrt{\sigma_{\rm fit,coadd}^{2} + \sigma_{\rm sys}^{2}}
    \label{eq:total_uncertainty}
\end{equation}

\noindent where $\sigma_{\rm fit,coadd}^{2}$ is the fit uncertainty as described in Section~\ref{sec:mc_uncertainties}, and $\sigma_{\rm sys}^{2}$ is the systematic uncertainty corresponding to the systematic uncertainty floor for either \feh\ or \alphafe\ as given above. 

\subsection{Comparison between coadded and weighted average abundances}
\label{sec:coadd_vs_weighted_average}
To test the accuracy of our pipeline, we compare the inverse variance weighted average \feh\ and \alphafe\ from measurements of individual spectra with that measured from coadded spectra. We would expect similar results if a coadded spectrum is truly representative of the weighted average of its constituents. 

\begin{figure*}
    \centering
	\includegraphics[width=0.95\textwidth]{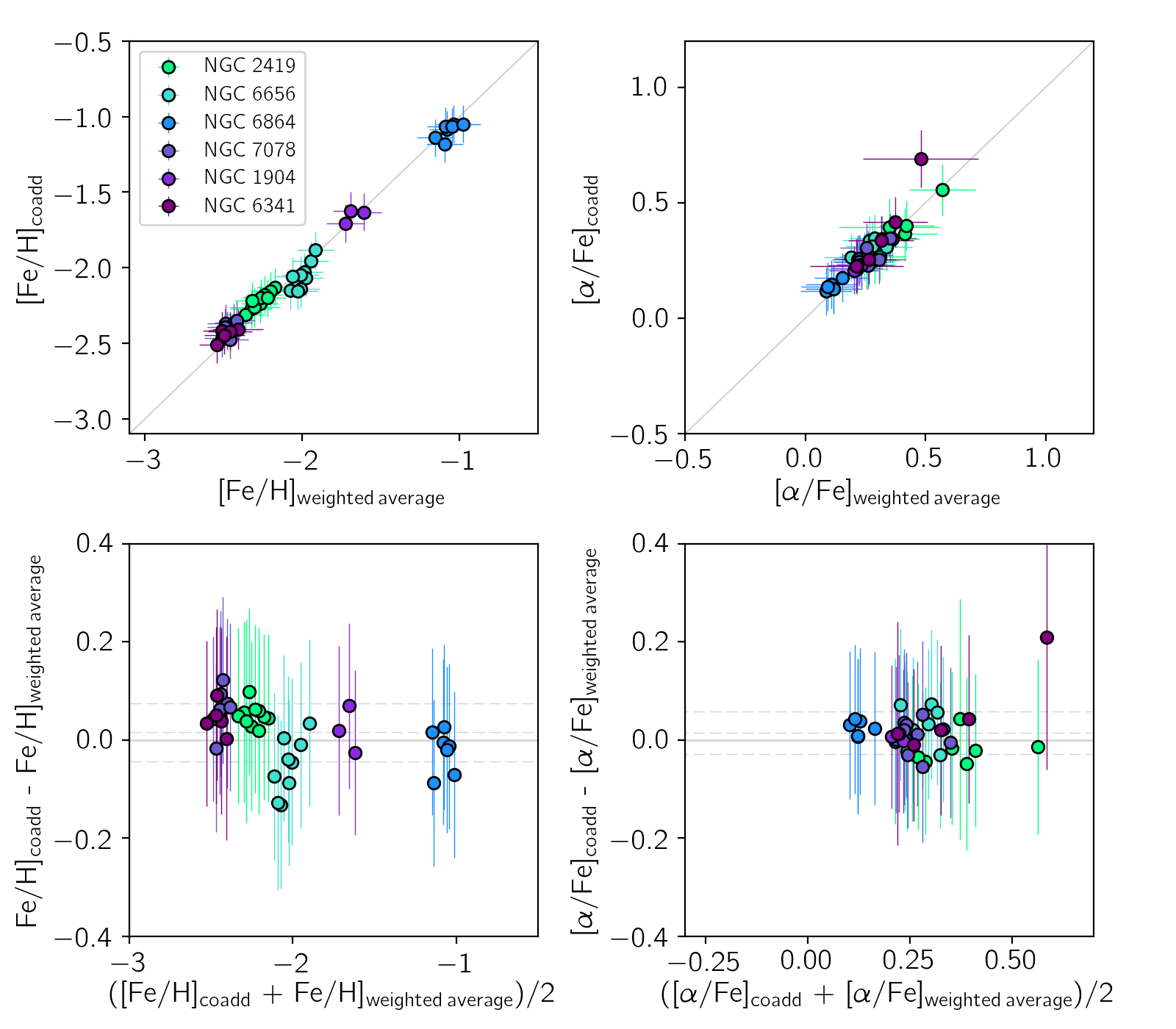}
    \caption{Top: Comparison between the weighted-average abundances with those measured from coadded spectra, for six Milky Way globular clusters.  Bottom left: residuals of the difference between the weighted-average abundances and those measured from coadded spectra. The dashed grey lines indicate the mean and standard deviation of the difference. The weighted average \feh\ and \alphafe\ values are determined as described in Section~\ref{sec:coadd_vs_weighted_average}. Values for the mean and dispersion of the residuals are given in the text. As the offset and dispersion are smaller than the average total uncertainty, we conclude that the measurement from our coaddition method accurately recovers the weighted average of abundance measurements from individual stars in a given coadd. }
    \label{fig:weighted_average_comparison_gcs}
\end{figure*}

\begin{figure*}
    \centering
	\includegraphics[width=0.95\textwidth]{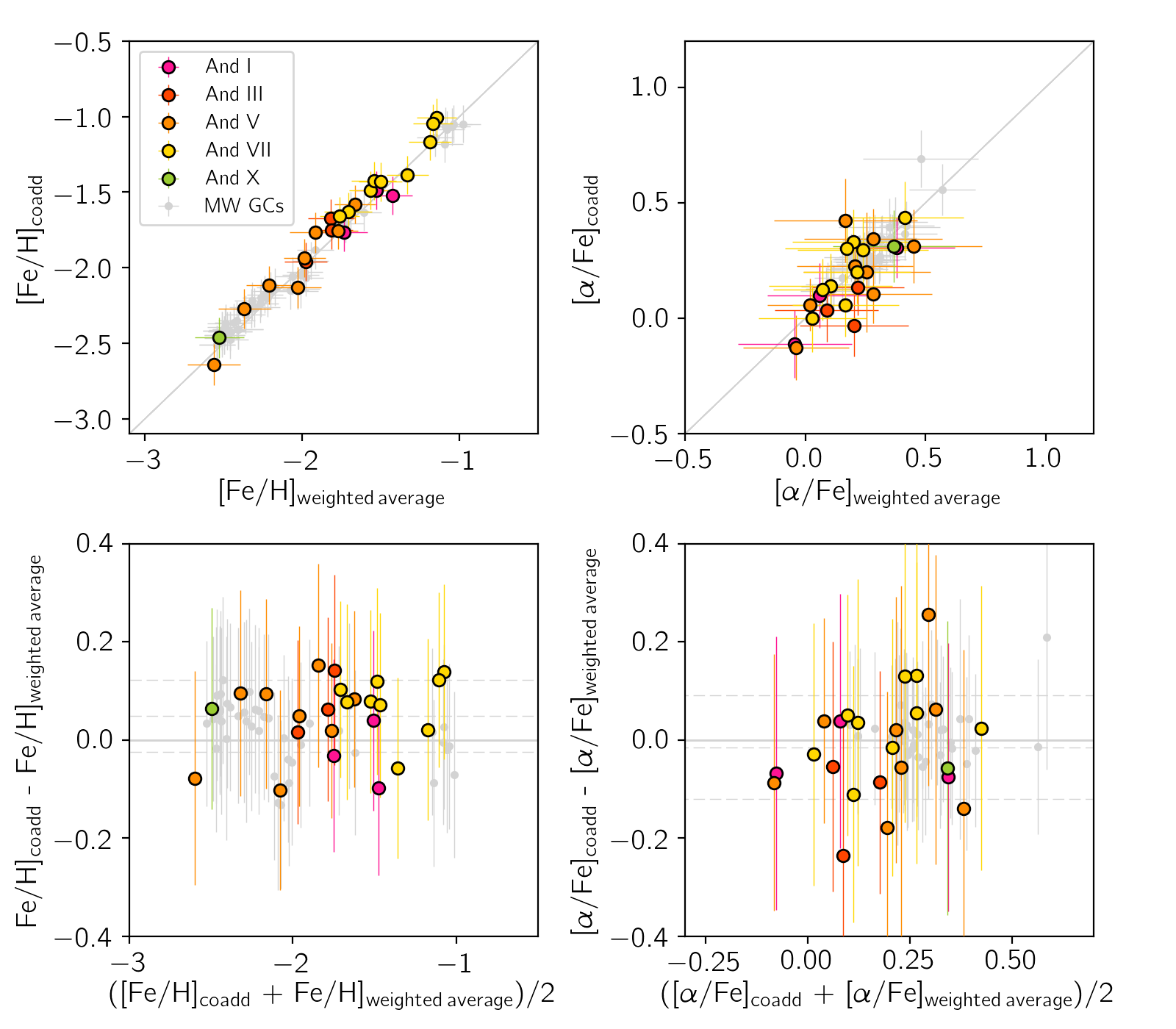}
    \caption{The same as Figure~\ref{fig:weighted_average_comparison_gcs}, with the colored points indicating our sample of validation coadds from M31 dSphs with deep observations ($\langle \mathrm{S/N}=18.2 \rangle$), and the grey points indicating the same MW GCs from Figure~\ref{fig:weighted_average_comparison_gcs}.}
    \label{fig:weighted_average_comparison_deep_dsphs}
\end{figure*}

\begin{figure*}
    \centering
	\includegraphics[width=0.95\textwidth]{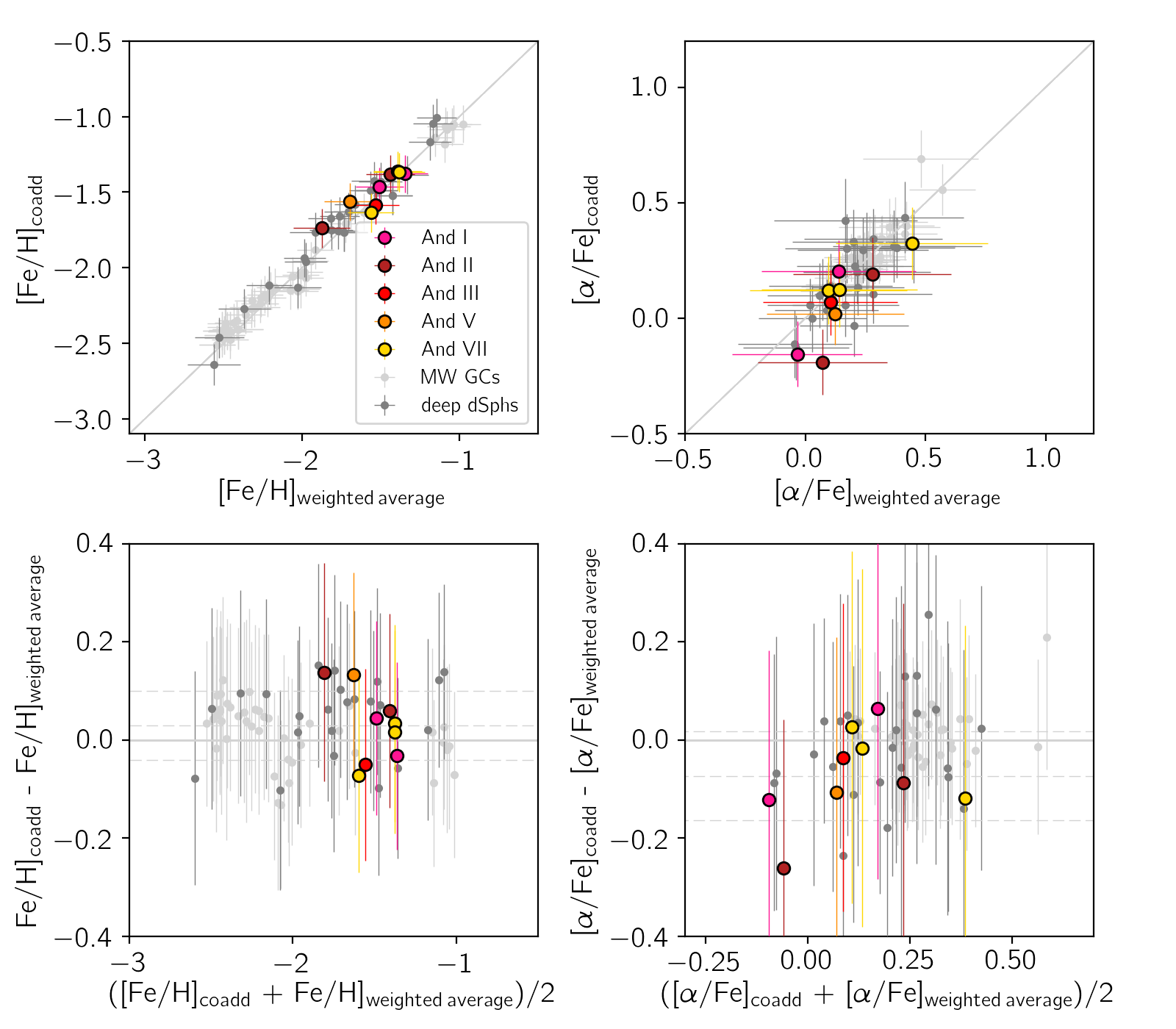}
    \caption{The same as Figure~\ref{fig:weighted_average_comparison_deep_dsphs}, with the colored points indicating our sample of M31 dSphs with shallow observations ($\langle \mathrm{S/N}=11.7 \rangle$), and the light and dark grey points indicating the same Milky Way globular clusters and M31 dSphs with deep observations from Figures~\ref{fig:weighted_average_comparison_gcs} and~\ref{fig:weighted_average_comparison_deep_dsphs}, respectively.}
    \label{fig:weighted_average_comparison_shallow_dsphs}
\end{figure*}

To compute the average inverse variance weighted abundance of a group of spectra, we define the following weights ($\omega_j$), following \citet{yang13}:

\begin{equation}
    \sigma^2_{\mathrm{spec},j} = \left(\sum_{i=1}^{\mathrm{mpixel}}(\sigma_{\mathrm{pixel},ij})^{-2}\cdot M_{\mathrm{element,}X}\right)^{-1}
\end{equation}

\begin{equation}
    \omega_j(X) = \frac{1/\sigma^2_{\mathrm{spec,j}}}{\sum_{j=1}^{n}1/\sigma^2_{\mathrm{spec,j}}}
\end{equation}
where mpixel is the total number of pixels in a spectrum, $\sigma_{\mathrm{pixel},ij}$ is the variance of the {\it i}th pixel of the {\it j}th spectrum in a group, and $M_{\mathrm{element,} X}$ is the corresponding elemental mask for fitting $X$, where $X$ is either [Fe/H] or [$\alpha$/Fe]. The weighted variance for the {\it j}th spectrum is $\sigma^2_{\mathrm{spec},j}$, and $\omega_j(X)$ is the weight of either [Fe/H] or [$\alpha$/Fe] for the {\it j}th star of $n$ stars in a coadded group. The weighted average abundance is then:

\begin{equation}
    X_{\mathrm{bin,wa}} = \sum_{j=1}^n \omega_j(X)X_j = \frac{\sum_{j=1}^{n}X_j/\sigma^2_{\mathrm{spec,j}}}{\sum_{j=1}^{n}1/\sigma^2_{\mathrm{spec,j}}}.
\end{equation}

The weighted uncertainty on the weighted average abundance is given as: 

\begin{equation}
    \sigma^2_{\mathrm{bin,wa}}(X) = \omega^2_j(X)\sigma^2_{\mathrm{total,j}}(X),
    \label{eq:total_uncertainty_wa}
\end{equation}

\begin{equation}
    \sigma_{\mathrm{total,j}}(X) = \sqrt{(\sigma_{\mathrm{fit,j}}(X))^2 + \sigma_{(\mathrm{sys}}(X))^2}
\end{equation}
\noindent where we adopt the systematic uncertainties of 0.101 dex for \feh, and 0.084 dex for \alphafe, from \citet{kirby20}, where the method used to measure these values is described by \citet{kirby10}.

In Figures~\ref{fig:weighted_average_comparison_gcs},  \ref{fig:weighted_average_comparison_deep_dsphs}, and \ref{fig:weighted_average_comparison_shallow_dsphs}, we compare the weighted average [Fe/H] and [$\alpha$/Fe] from measurements of individual spectra with their measurements from coadds for MW GCs, M31 dSphs with deep observations, and M31 dSphs with shallow observations, respectively. In each consecutive plot, we include the points from the previous plot (as grey points), for reference and to illustrate the range in \feh\ and \alphafe\ that each sample covers.

\subsubsection{Milky Way globular clusters}
\label{sec:mw_gcs_coadd_wa_comparison}
Our MW GC sample includes stars that are much closer (average distance 23\,kpc), and are therefore much brighter (average $\sim 4$ mag brighter in $V$-band) than RGB stars in M31 dSphs. For these GC stars, we obtain high S/N spectra, which in turn provide a higher degree of certainty on their individual chemical abundance measurements. Because GCs are principally single stellar populations, we also can safely assume that the metallicity (\feh) distribution function of these GCs is roughly single-valued, and can therefore more clearly identify outliers in \feh. In  Figure~\ref{fig:weighted_average_comparison_gcs}, we show the comparison between the weighted average \feh\ and \alphafe\ from individual stars to those measured from the coadds. From the residuals (bottom row of Figure~\ref{fig:weighted_average_comparison_gcs}), we find an average offset of $-0.008$ dex in \feh\ with a dispersion of 0.057 dex. We consider this to be very good agreement between the weighted-average and coadded abundance measurements, across the large range of [Fe/H] that our GCs probe. In general, we do not find any trend in the difference as a function of metallicity. For \alphafe, we find an average difference of 0.027 dex, with a dispersion of 0.033 dex. Again, we consider this to be very good agreement between the weighted average and coadded \alphafe\ measurements, as we find no significant offset in the difference, and the dispersion is less than the adopted systematic uncertainty. 

\subsubsection{M31 dSphs with deep observations}
\label{sec:m31_dsphs_deep_coadd_wa_comparison}
Figure~\ref{fig:weighted_average_comparison_deep_dsphs} shows the same comparison as Figure~\ref{fig:weighted_average_comparison_gcs}, but now with the colored points showing our groupings for M31 dSph stars with deep observations ($\langle \mathrm{S/N}=18.2 \rangle$, see Figure~\ref{fig:sample_snr_hist}). In this figure we also include MW GC groupings, for comparison, in grey. For \feh, we measure an average difference of 0.042 dex, with a dispersion of 0.076 dex, and for \alphafe, we measure an average offset of $-0.007$ dex with a dispersion of 0.099 dex. While the spread in differences is larger for this sample, any offset and associated dispersion is small compared to the total uncertainty for the coadded measurements. For both \feh\ and \alphafe, we do not find any trend in the difference as a function of \feh\ or \alphafe.

\subsubsection{M31 dSphs with shallow observations}
\label{sec:m31_dsphs_shallow_coadd_wa_comparison}
In Figure~\ref{fig:weighted_average_comparison_shallow_dsphs} we consider our sample of M31 dSphs for which we have shallow observations (colored points), with MW GCs and deep dSph samples for comparison in light grey and grey, respectively. This sample contains, on average, stars with much lower S/N spectra ($\langle \mathrm{S/N}=11.7 \rangle$, see Figure~\ref{fig:sample_snr_hist}). For \feh, we find an offset of 0.022 dex with a dispersion of 0.066 dex, and for \alphafe, we find an offset of $-0.053$ dex, with a dispersion of 0.074 dex, and no trend in difference with measured abundance for either \feh\ or \alphafe. We find, in general, that the difference between the weighted-average and coadded abundances for shallow M31 dSph spectra is consistent with those from deep observations, illustrating the performance of our method at the lower end of the S/N regime.

If we consider all three samples simultaneously, we find an average offset in \feh\ of 0.015 dex, with a dispersion of 0.074 dex, and for \alphafe, we find an offset of 0.002 with a dispersion of 0.076 dex. Unlike \citet{yang13}, we do not find any trend in average difference as a function of \feh\ or \alphafe\ for any of our samples (see their Figure~8). These offsets are approximately an order of magnitude lower than the systematic uncertainty floors (0.101 for \feh, 0.084 for \alphafe). As a result, we conclude that our coaddition method accurately reproduces the same \feh\ or \alphafe\ measurement as the weighted average of the individual spectra in a group. 

\section{Coadding spectra with large uncertainties on their individual abundance measurements}
\label{sec:coadd}
So far, we have only considered stars for which we have secure abundance measurements, i.e., uncertainties in [Fe/H] and [$\alpha$/Fe] less than 0.4 dex, and the $\chi^2$ minimum is not at the edge of the grid of synthetic spectra ($[-5,0.0]$ for [Fe/H], $[-0.8,1.2]$ for [$\alpha$/Fe]). However, our aim is to test the application of our method to stars in M31 where we cannot reliably measure abundances individually (most of the spectra in SPLASH). Therefore, for the following analysis, we consider only stars from the M31 dSph slitmasks. In addition, for the purposes of investigating the abundance distributions, we no longer consider deep and shallow coadds separately, combining both sets of coadds to obtain the abundance distribution for a given slitmask. We will refer to the coadds used for validation as our ``validation coadds", and the coadds that were not used for validation as ``science coadds". Our coadded abundance measurements for both \feh\ and \alphafe\ for both validation and science coadds can be found in Table~\ref{tab:coadded_abundances}. 

\begin{deluxetable*}{llllllrl}
\tablenum{4}
\tablecaption{Excerpt of measured \feh\ and \alphafe\ from our coadded spectra. Includes both validation and science coadds, for both bin types.\label{tab:coadded_abundances}}
\tablehead{\colhead{Bin type} & \colhead{Coadd type} & \colhead{Mask name} & \colhead{Slit list} & \colhead{\feh} & \colhead{$\sigma$\feh} & \colhead{\alphafe} &  \colhead{$\sigma$\alphafe}}
\startdata
\feh&validation&and7a&[117, 063, 094, 004, 108]&$-1.63$&0.11&0.30&0.13\\
\feh&validation&and7a&[007, 076, 083, 071, 068]&$-1.43$&0.10&0.06&0.12\\
\feh&validation&and7a&[124, 000, 062, 067, 065]&$-1.17$&0.10&0.44&0.14\\
\feh&validation&and7a&[047, 123, 075, 049, 026]&$-1.39$&0.11&0.33&0.13\\
\feh&validation&and7a&[034, 020, 011, 090, 058]&$-1.43$&0.11&0.30&0.14\\
\feh&validation&and7a&[073, 091, 060, 125, 061]&$-1.66$&0.10&0.12&0.12\\
\feh&science&and7a&[040, 010, 029, 059, 015, 084, 054, 112, 107]&$-1.77$&0.11&$-0.26$&0.16\\
\feh&science&and7a&[066, 113, 120, 086, 081]&$-1.70$&0.12&0.40&0.20\\
\feh&science&and7a&[074, 019, 023, 082, 069]&$-1.57$&0.14&$-0.10$&0.33\\
\feh&science&and7a&[043, 110, 111, 085, 016]&$-1.49$&0.12&0.18&0.20\\
\feh&science&and7a&[051, 018, 025, 012, 024]&$-1.57$&0.11&$-0.35$&0.18\\
\hline
\enddata
\tablecomments{This table is available in its entirety in machine-readable form.}
\end{deluxetable*}

We apply the same M31 dSph membership criteria as described in Section~\ref{sec:membership_quality_criteria}, now coadding only those spectra where the individual abundance measurements did not satisfy the quality criteria. We now also consider additional slitmasks for which we only have shallow ($\sim 1-2$ hr) observations, adding And IX, XIV, XV, and XVIII to our sample of M31 dSphs. This increases our sample size to ten total satellite galaxies. We then apply the same coaddition method as described in Section~\ref{sec:coadded_abundances}. Table~\ref{tab:failed_groupings} gives a summary of the science coadds for each slitmask.

For these science coadds, we cannot directly compare abundance measurements from the coadded spectra with those from the weighted average, as they do not have reliable individual \feh\ and/or \alphafe. Instead, we compare the distribution of these measurements with that of the validation sample, i.e., where latter consists of the individual stars having reliable measurements. 

\begin{figure*}
    \centering
	\includegraphics[width=0.8\textwidth]{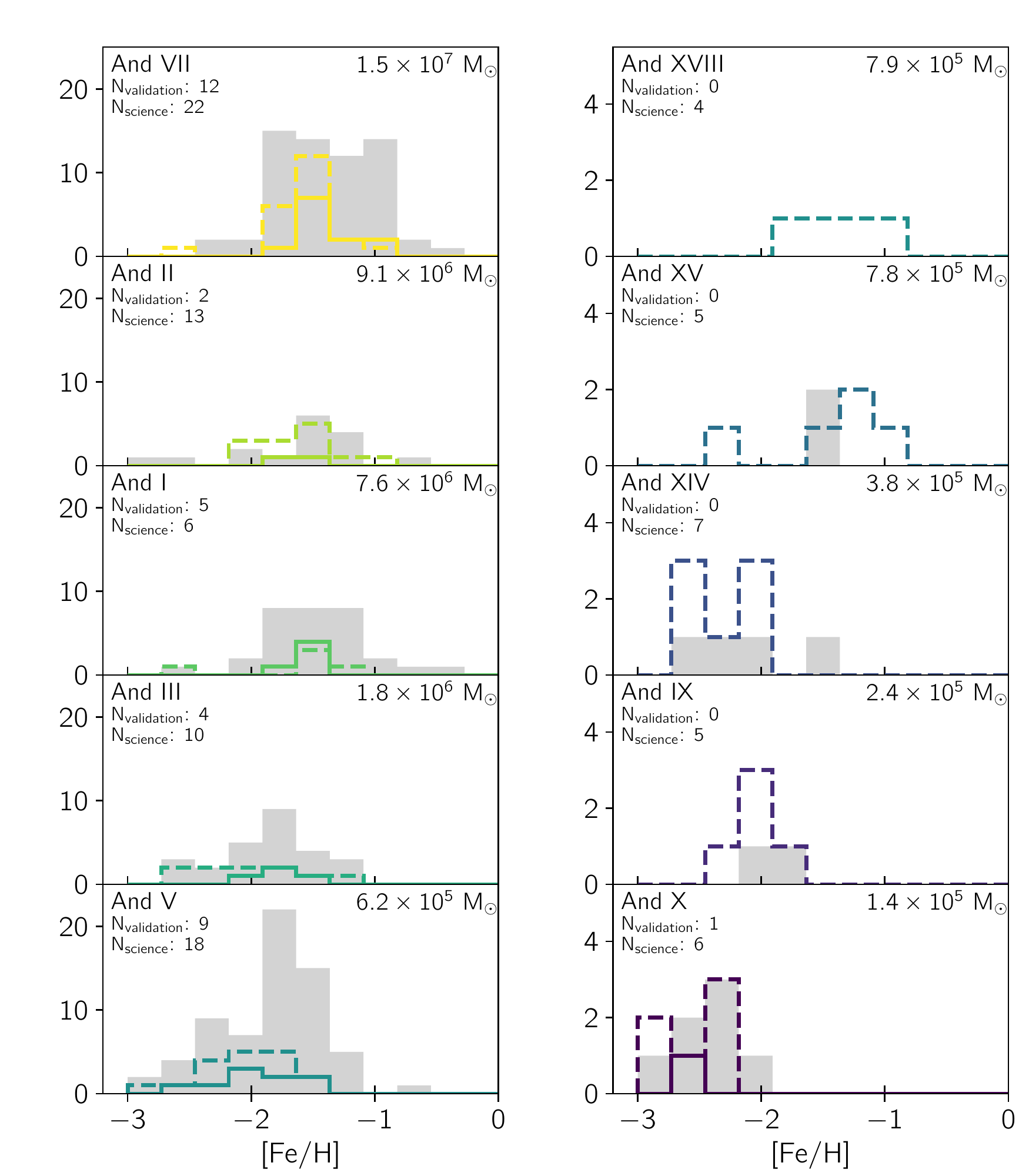}
    \caption{[Fe/H] distributions for our ten M31 dSphs roughly ordered from most massive to least massive (left column, top to bottom, then right column, top to bottom). The filled grey histogram represents the individual \feh\ measurements from both shallow and deep (low and high S/N) data. The colored histograms represent the measurements made from coadded spectra, where the solid line denotes the validation coadds, and the dashed line indicates the science coadds. The stellar mass of each dSph is indicated in its respective panel.}
    \label{fig:good_vs_failed_feh_hist}
\end{figure*}

In Figure~\ref{fig:good_vs_failed_feh_hist}, we show histograms of spectroscopic \feh\ measurements for three samples: individual measurements (solid grey), validation coadds (solid colored lines), and science coadds (dashed colored lines). The ten M31 dSphs in our sample are sorted roughly by their stellar mass. We note that And V is out of order with respect to mass, but we keep it in the left-hand column as the sample size is more akin to those in the left-hand column. 

We use a set of statistical tests to determine how the \feh\ distribution of our science coadds compares to the underlying distribution of individual stars that they should represent. First, we calculate the Kolmogorov-Smirnov statistic between the distribution of individual stars and both the validation coadds, and between the individual stars and science coadds. For dSphs that have stars with both individual abundance measurements and validation coadds, we find that the K-S test cannot reject the null hypothesis that the two distributions are the same. Some caution is warranted for the dSphs for which we have only a few individual or coadded measurements (And XVIII, XV, XIV, IX, X), as this test may be less reliable, due to the sparse sampling of the abundance distributions.

When we compare abundance measurements from individual stars to our science coadds, we find that, for three dSphs in our sample (And VII, III, and V),  the null hypothesis can be rejected, i.e., it is more likely that the samples are not drawn from the same distribution. For these three dSphs, we also perform Welch's t-test between the distribution of validation and science coadds, to test if the two distributions have similar means, though not necessarily similar population variances. With this test, we find that the distributions of the validation and science coadds agree for And III and V, but the null hypothesis is rejected for And VII\@. We find that while the distribution of science coadds fits within the distribution of individual abundance measurements for the majority of our M31 dSphs, there is a slight shift in the distribution mean between the validation coadds and the science coadds. The distributions of science coadds tend to be more metal-poor than the validation coadds. We note that this is somewhat expected, as it is more difficult to determine abundances for an individual metal-poor star compared to a metal-rich star in this low S/N regime, owing to the depth of the spectral features. Therefore, these stars may be less likely to be part of the validation sample, and more likely to be used for the science coadds.

In general, we find that our science coadds agree with the distribution from our validation sample, and fit within the distribution of individual abundance measurements. We further expand on this comparison by exploring the 2D \feh--\alphafe\ chemical space in the following section.

\begin{table}
    \renewcommand\thetable{5}
	\centering
	\caption{Coadds from stars with large uncertainties on their individual abundance measurements (science coadds).}
	\label{tab:failed_groupings}
    \begin{tabular}{llll}
	\hline
	\hline
Object name & Mask name & N$_{\rm stars}$ & N$_{\rm science\,coadds}$ \\
    \hline
And I       & and1a     & 10          & 2                        \\
            & d1\_1     & 10          & 2                        \\
            & d1\_2     & 14          & 2                        \\
And II      & d2\_1     & 37          & 7                        \\
            & d2\_2     & 34          & 6                        \\
And III     & and3a     & 11          & 2                        \\
            & d3\_1     & 20          & 4                        \\
            & d3\_2     & 12          & 2                        \\
            & d3\_3     & 10          & 2                        \\
And V       & and5a     & 18          & 3                        \\
            & and5b     & 22          & 4                        \\
            & d5\_1     & 20          & 4                        \\
            & d5\_2     & 24          & 4                        \\
            & d5\_3     & 18          & 3                        \\
And VII     & and7a     & 29          & 5                        \\
            & d7\_1     & 50          & 10                       \\
            & d7\_2     & 36          & 7                        \\
            & d7\_3     & 17          & 3                        \\
And IX      & d9\_1     & 19          & 3                        \\
            & d9\_2     & 10          & 2                        \\
And X       & and10a    & 10          & 2                        \\
            & and10b    & 6           & 1                        \\
            & d10\_1    & 7           & 1                        \\
            & d10\_2    & 11          & 2                        \\
And XIV     & A170\_1   & 12          & 2                        \\
            & A170\_2   & 24          & 4                        \\
            & d14\_3    & 8           & 1                        \\
And XV      & d15\_1    & 16          & 3                        \\
            & d15\_2    & 10          & 2                        \\
And XVIII   & d18\_1    & 20          & 4                       \\
	\hline
	\end{tabular}
\end{table}

\section{Discussion and applications}
\label{sec:alphafe_feh}
\subsection{The \alphafe--\feh\ plane}
In Figure~\ref{fig:feh_alphafe_dsphs}, we present the  \alphafe--\feh\ distributions for the ten M31 dSphs in our sample, with measurements from individual stars indicated by transparent grey circles, measurements from validation coadds with solid circles, and measurements from science coadds with diamonds. Measurements made from shallow data are color-coded according to a galaxy's stellar mass, and measurements made from deep data are shown in grey. The dSphs are sorted approximately by mass, so that the relationship of decreasing average metallicity with decreasing mass is easily visible. As in Figure~\ref{fig:good_vs_failed_feh_hist}, we keep And V in the left-hand column, as it more relevant to compare it to the more massive satellites with a larger sample size. The less-massive dSphs are more sparsely sampled, but we find they still follow roughly the same trend (see Section~\ref{sec:MZR} for an in-depth discussion). We note that for four dSphs in our sample (And XVIII, XV, XIV, and IX), our \alphafe\ measurements are the first available for these dwarf galaxies. 

For the majority of the dSphs in our sample, we also see a clear decrease in \alphafe\ as a function of \feh\ for individual stars (transparent grey points). Even where we have a significantly smaller sample of stars (e.g., And XIV, And X), there is still a visibly negative trend, which is further supported by the inclusion of our coadded abundances. We quantify these trends in Section~\ref{sec:sfh_of_dsphs}. The validation coadd abundance measurements (circles) generally fall neatly within the center of the 2D \alphafe--\feh\ distribution of the individual stellar abundances, while the science coadded measurements have a wider spread in both \feh\ and \alphafe. This echoes what we have already found in Figure~\ref{fig:good_vs_failed_feh_hist} and with the statistical tests performed in Section~\ref{sec:coadd}: the distributions of the validation sample and coadded measurements for a given dSph generally have the same mean, but differing variance. 

\citet{vargas14a} measured \alphafe\ and \feh\ for individual stars in six of the dSphs in our sample (And I, II, III, V, VII, X), and so here we qualitatively compare our results with those presented in their Figure 4. We note that they did not cull their sample to remove stars with large uncertainties, where we have an uncertainty cut at 0.4 dex in both \alphafe\ and \feh\ for individual stars in our sample. In addition, they required that spectra in their sample have a minimum S/N of 15 \AA$^{-1}$, while we have no minimum S/N requirement for stars in our sample. Finally, their sample consists of stars from slitmasks with shallow observations only. We find general agreement in the \alphafe--\feh\ distributions for the dSphs in common between our samples, with one notable exception. \citet{vargas14a} found a significant plateau in \alphafe\ for the metal-rich (\feh $> \sim -1.5$ dex) stars in And VII. In contrast, we find that while we have a number of high-\alphafe, metal-rich stars in And VII, they do not form a plateau. However, we note that while we measure a negative trend in the individual stars (Section~\ref{sec:sfh_of_dsphs}), we do not measure a significant negative trend with our coadded measurements. 

\subsection{Star formation history of M31 dSphs from coadded spectra}\label{sec:sfh_of_dsphs}

In Section~\ref{sec:alphafe_feh}, we presented \feh\ and \alphafe\ for ten dSphs in M31, nine of which have at least one measurement from individual stars, and all have coadded measurements that are representative of the distribution of individual abundance measurements. Five of these dSphs were also presented by \citet{kirby20}: And VII, And I, And III, And V, and And X, where in addition to atmospheric \alphafe, they also measure individual $\alpha$-elements (Mg, Ca, Si, Ti) for a number of stars in their sample with sufficiently high S/N. In this section, we will focus on the star formation history (SFH) from coadded measurements in M31 dSphs at the lower end of the mass range of our sample (And V - And X), as significant discussion of SFH of the higher mass dSphs (And VII, II, I, and III) based on this chemical space were discussed at length in previous works \citep[e.g.][]{kirby20,vargas14a}. 

\citet{vargas14a} used the \alphafe--\feh\ plane to characterize the SFHs of the dSphs in their sample, and compared their results to those in the literature from photometric studies of their CMDs  \citep[e.g.,][]{grebel99,dacosta00,dacosta02,mcconnachie07,weisz14a}. To summarize the findings from photometric studies: the majority of M31 dSphs show evidence for an extended SFH from their CMDs, with prominent red horizontal branch (HB) stars. The presence of a red HB indicates stars that formed more recently at multiple metallicities, compared to the primarily metal-poor, ancient stars that populate the blue HB. \citet{martin17} reached similar conclusions for some of the less massive satellites (And XVIII, XV, IX, X) in our sample. 

\begin{figure*}
    \centering
	\includegraphics[width=0.9\textwidth]{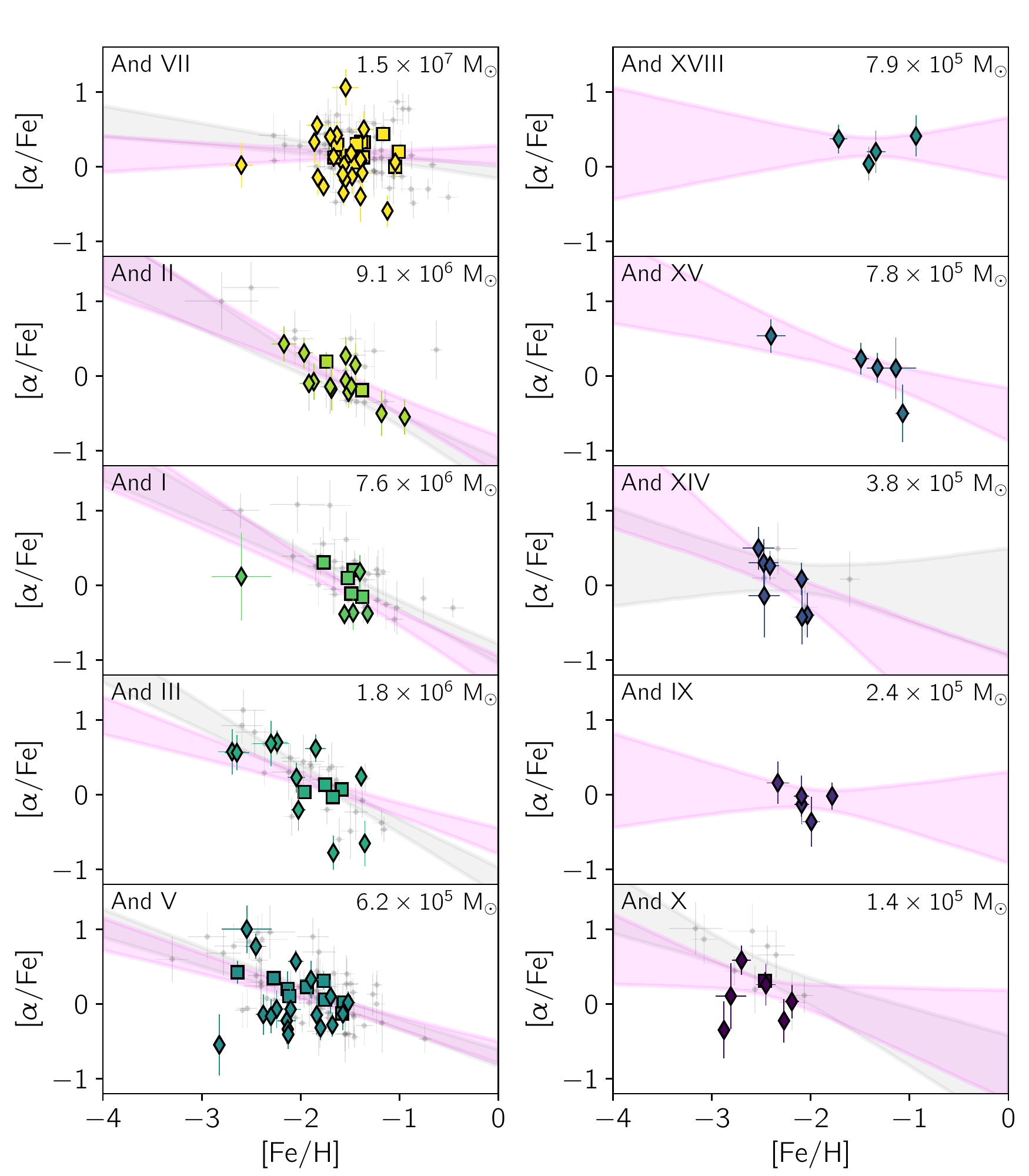}
    \caption{[Fe/H] vs [$\alpha$/Fe] for our ten M31 dSphs, ordered from most massive to least massive. Small grey circles indicate abundance measurements of  individual stars. Square symbols show validation coadds, and diamond symbols indicate science coadds. Fitted slopes and their corresponding $1\sigma$ uncertainties measured from individual measurements are shown in grey, and the slopes from all coadds (validation and science combined) for a given dSph are shown in magenta.
    }
    \label{fig:feh_alphafe_dsphs}
\end{figure*}

 We find that our coadded measurements follow the same decreasing \alphafe\ with \feh\ trend seen in individual stars. The more massive satellites in our sample (left column of Figure~\ref{fig:feh_alphafe_dsphs}) illustrate this trend clearly, while the less massive satellites generally exhibit negative trends, albeit with smaller sample sizes. In general, a decreasing trend in \alphafe\ as a function of \feh\ indicates an extended SFH, which is consistent with what has been inferred from the presence of red HB stars in their CMDs.

We quantitatively investigate the SFHs from coadded measurements in our dSphs by measuring the slope in \alphafe--\feh\ space ($d$\alphafe/$d$\feh), where \alphafe\ acts as a proxy for stellar age, and \feh\ provides a measure of the chemical enrichment. To measure $d$\alphafe/$d$\feh, we use a marginalized posterior distribution with flat priors to fit a line to data with uncertainties on both the dependent and independent axes, presented by \citet{hogg10} and described in detail for a similar application in \citet{kirby19ni}. 

We use a Markov Chain Monte Carlo (MCMC), implemented with the emcee Python package \citep{foreman-mackey13}, to generate a distribution of models which fit our data while also taking uncertainties into account. Rather than fitting a line with a slope ($m$) and intercept ($b$), we parameterize the line of best fit with an angle ($\theta$), and the perpendicular distance of the line from the origin ($b_{\perp}$), where $b_{\perp} \equiv b\, cos(\theta)$ \citep{hogg10}. This parameterization avoids the preference for shallow slopes when there is a flat prior on the slope. The log likelihood is defined as: 

\begin{equation}
   \mathrm{ln}\;\mathscr{L} = -0.5 \sum_{i = 1}^{N}\frac{\Delta^2_i}{\Sigma^2_i} - \mathrm{ln}(\sqrt{2\pi}\Sigma_i)
\end{equation}

\begin{equation}
   \Delta_i = [\alpha/\mathrm{Fe}]_i\,\mathrm{cos}\theta - \mathrm{[Fe/H]}_i\,\mathrm{sin}\theta - b_{\perp}
\end{equation}

\begin{equation}
   \Sigma_i^2 = \delta[\alpha/\mathrm{Fe}]_i^2\,\mathrm{cos}^2\theta + \delta\mathrm{[Fe/H]}_i^2\,\mathrm{sin}^2\theta
\end{equation}

\noindent where for star $i$, $\delta$[Fe/H]$_i$ and $\delta$[$\alpha$/Fe]$_i$ represent the total measurement uncertainties for \feh\ and \alphafe, respectively. Using the emcee package, we sample the probability distribution, and transform our parameters back into the more familiar $m$ and $b$ using $m=\mathrm{tan}(\theta)$ and $b = b_{\perp}/\mathrm{cos}(\theta)$. For both $\theta$ and $b_{\perp}$ we take the median of the resulting  distribution as our measured value, and the 16th and 84th percentiles ($1\sigma$) as the uncertainties on our measured values. 

We measure slopes for both the individual measurements (transparent grey points in Figure~\ref{fig:feh_alphafe_dsphs}), as well as the slope for all validation and science coadds combined. The measured slopes and their associated uncertainties are given in Table~\ref{tab:alphafe_feh_slopes}. We find that for the majority of our sample, the slopes from measured individual abundance measurements and coadds roughly agree, within $\sim1$~-~$2\sigma$. We interpret this finding as another indication that, in general, abundance measurements from our coadded spectra are representative of the overall abundance measurements from individual spectra. 

We find evidence for slopes in \alphafe$-$\feh\ from our coadded measurements at the 2$\sigma$ significance level for all of our low-mass galaxies with the exception of And VII, And IX, and And XVIII. For And VII, we interpret this as a consequence of the presence of metal-rich, high-\alphafe\ stars, which would flatten a negative trend in \alphafe--\feh. Such metal-rich, high-\alphafe\ stars are absent from our And IX sample, and we discuss below potential reasons for a lack of a significant trend for this dSph. Finally, we note that the slope for And XVIII was measured using only four coadded points. 



For our stars in And VII with individual measurements, we find a negative slope in $d$\alphafe/$d$\feh\ that roughly corresponds with that found by \citet{kirby20}, and this negative slope is significantly more shallow compared to other galaxies in the sample. Moreover, using our coadded measurements, we do not find evidence for a negative correlation. These findings, along with the presence of high-\alphafe, metal-rich stars is in contrast to the SFH from shallow HST photometry \citep{weisz14a}, where they find And VII to have an ancient SFH\@. To resolve this disparity, deeper spectroscopic observations of And VII are needed, as noted by \citet{kirby20}. 

The slopes that we measure quantitatively agree with \citet{kirby20}, where the shallower slopes of the more massive galaxies indicate higher star formation rates over an extended period. They compare the observed metallicity distribution function (MDF) with chemical evolution models \citep{kirby11,kirby13} to find a best fit chemical evolution model, and compare these best fit models with the slopes they measure in \alphafe--\feh. They find that galaxies with shallower slopes are incompatible with a simple Leaky Box Model \citep{schmidt63,talbot71}, and are more suggestive of the Pre-Enriched Model \citep{pagel97}, or a model where gas accretion prolongs star formation \citep[Accretion Model, ][]{lynden-bell75}. In the Leaky Box model, a galaxy can lose gas at the same rate of its star formation, but does not acquire any new gas. The fact that we find more massive galaxies to have shallower slopes in \alphafe--\feh\ indicates they are more likely to hold on to their gas, or have had some accretion of gas, likely owing to the depth of their gravitational potential wells (in comparison to the less massive dSphs) \citep{Tolstoy09,letarte10,kirby11}. As a result, these galaxies have a less efficient SFH, where Type Ia supernovae, which produce more Fe than $\alpha$ elements, dominate over Type II supernovae. 

We take the presence of a slope in our lower mass galaxies as further evidence of an extended star formation history in these galaxies, as also evidenced by the presence of red HB stars in the CMDs. And XIV in particular is interesting because it hosts a more balanced (i.e., a more equal number of red HB and blue HB stars) HB \citep{martin17} in comparison to the other dSphs in the low-mass regime. From our spectroscopic abundance measurements, we find that it is more metal-poor than one would naively expect following the mass-metallicity relation (MZR, Section~\ref{sec:MZR}). These  spectroscopic measurements corroborate the evidence from photometric data that And XIV hosts a more substantial fraction of old, metal-poor stars in comparison to other dSphs of similar mass. 

We also find tentative evidence for a slope in the \alphafe--\feh\ distribution for And IX ($m = -0.12\pm0.10$). 
Using HST photometry, \citet{martin17} found And IX was relatively metal-rich for its stellar mass, but they noted that their sample likely had some contamination from the M31 halo. Such contamination would affect the slope by making it more flat. However, even considering our strict membership criteria, we still find that And IX is more metal-rich than other dSphs of similar mass. We note that the slope ($d$\alphafe/$d$\feh) we measure from coadded stars in And IX is significantly shallower compared to other dSphs with similar stellar masses. This difference in the slope may be due to the presence of metal-rich stars (which are undetected in our sample of And IX stars, unlike for And VII), large uncertainties in \alphafe, or simply a lack of spectroscopic metallicity measurements at the metal-poor end of the distribution. As the slope from coadded measurements is calculated from only five stars, we refrain from drawing further conclusions.


\subsection{The mass-metallicity relation}
\label{sec:MZR}
\citet{kirby13} presented a mass-metallicity relation (MZR) for dwarf galaxies in the Local Group, fit using 15 MW dSphs. In their analysis, they found from coadded spectra that the M31 dSphs in their sample roughly corresponded to the same relation, indicating that satellite galaxies of the MW and M31 share the same MZR, which is roughly consistent with the MZR measured from more massive galaxies. However, for the majority of their M31 dSphs, the average metallicity was calculated by coadding stars in only a few bins, and in some cases, only one bin was used for a given dSph. \citet{kirby20} update this result by remeasuring the mean \feh\ from measurements from individual stars for five M31 dSphs with deep spectroscopic observations (And I, III, V, VII, X). They find that their new values agree within 0.2 dex for all dSphs in their sample except for And VII, which agrees within 0.25 dex. 

We update this finding by combining both the individual measurements and the coadded measurements for \feh for each dSph. To do this, we take the uncertainty-weighted average of the individual and coadded measurements, where the total uncertainty for individual stars is given in Section~\ref{sec:pipeline_process}, and the total uncertainty for our coadded measurements is given by Eq.~\ref{eq:total_uncertainty}. Stellar mass and mean \feh\ are given in Table~\ref{tab:alphafe_feh_slopes} for each dSph in our sample. {Stellar masses for M31 dSphs are the same as in  \citet{kirby20} and \citet{kirby13}, where the mass-to-light ratios from \citet{woo08} are multiplied by luminosities presented in \citet{tollerud12} and \citet{mcconnachie12}.}

\begin{figure}
	\includegraphics[width=\columnwidth]{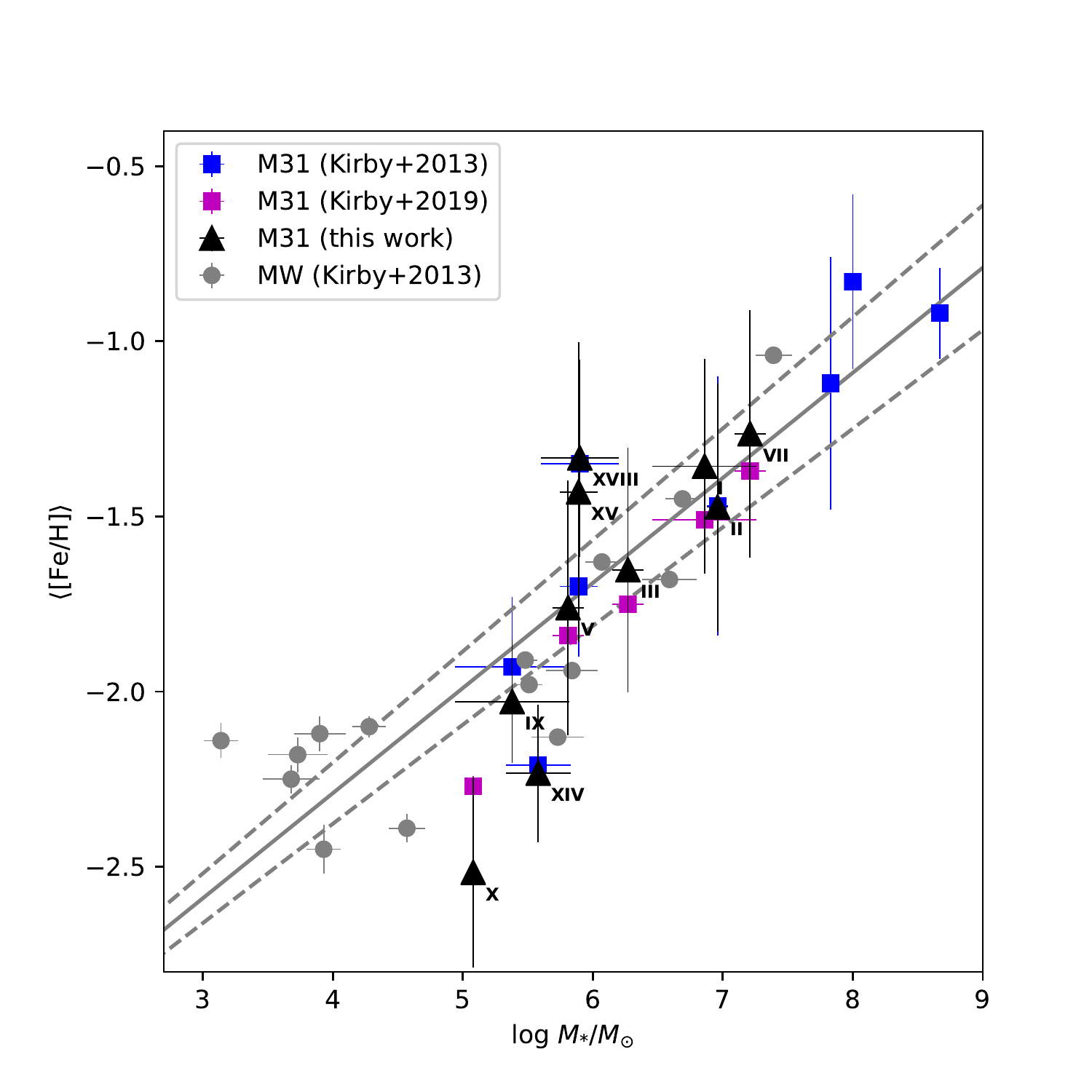}
    \caption{Mass-metallicity relation for Local Group satellites. Average metallicities measured with individual stars in MW dSphs from \citet{kirby13} are indicated with grey circles, average metallicities measured from coadded spectra for M31 dSphs from \citet{kirby13} are indicated with blue squares, and average metallicities measured from individual stars in M31 dSphs from \citet{kirby20} are indicated with purple squares. Our measurements are indicated with black triangles, labelled with the name of the M31 dSph. The solid grey line shows the MZR from \citet{kirby13}, {with the dashed lines indicating $1\sigma$ uncertainties}, fit using MW dSphs only. The average metallicity of M31 dSphs, whether measured from individual or coadded spectra, fit this relation.}
    \label{fig:mass_metallicity}
\end{figure}

Figure~\ref{fig:mass_metallicity} shows the MZR from our sample of M31 dSphs, compared to the values from \citet{kirby13} and \citet{kirby20}. Overall, we find that our measurements (black triangles) are consistent with the mean \feh\ from either coadded measurements \citep{kirby13} or individual measurements \citep{kirby20}. {We note that while And X and XIV seem to deviate from the MZR fit from MW dSphs, our measurements are consistent within $\sim1-2 \sigma$ uncertainty.} Using our updated sample, which includes coadded abundances from stars that did not have high enough S/N to derive individual measurements, we demonstrate that we can robustly determine the average metallicity of a given M31 dSph by properly accounting for these stars. 

\begin{deluxetable*}{lllll}
\tablenum{6}
\tablecaption{Mass-metallicity Relation and Abundance Trends in M31 dSphs.\label{tab:alphafe_feh_slopes}}
\tablehead{\colhead{Object name} & \colhead{log(M$_*$/M$_{\sun}$)} & \colhead{$\langle$[Fe/H]$\rangle$}  & \colhead{m (individual)} & \colhead{m (coadd)}}
\startdata
And VII & $7.21\pm0.12$* & $-1.26\pm0.003$& $-0.17\pm0.07$ & $+0.01\pm0.09$ \\
And II & $6.96\pm0.08$ & $-1.47\pm0.006$&  $-0.72\pm0.14$ & $-0.64\pm0.16$ \\
And I & $6.86\pm0.40$* & $-1.36\pm0.004$& $-0.65\pm0.10$ & $-0.73\pm0.15$ \\
And III & $6.27\pm0.12$* & $-1.65\pm0.005$& $-0.73\pm0.11$ & $-0.42\pm0.10$ \\
And V & $5.81\pm0.12$* & $-1.76\pm0.003$ &$-0.45\pm0.06$ & $-0.40\pm0.08$ \\
And XVIII & $5.90\pm0.30$ & $-1.33\pm0.02$& ... &  $-0.01\pm0.25$ \\
And XV & $5.89\pm0.145$ & $-1.43\pm0.42$ & ... & $-0.43\pm0.21$ \\
And XIV & $5.58\pm0.25$  & $-2.23\pm0.01$ & $-0.16\pm0.29$ & $-0.80\pm0.82$\\
And IX & $5.38\pm0.44$ & $-2.03\pm0.01$& ... & $-0.12\pm0.26$ \\
And X & $5.08\pm0.03$* & $-2.52\pm0.27$ & $-0.60\pm0.21$ & $-0.32\pm0.25$ \\
\hline
\enddata
\tablecomments{Masses indicated with asterisks are from \citet{kirby20}, and are the product of the stellar mass-to-light ratios from \citet{woo08} and luminosities from \citet{tollerud12}. Masses without asterisks are from \citep{kirby13} and references therein.}
\end{deluxetable*}

\section{Conclusions}
\label{sec:conclusions}
We present a coaddition method for DEIMOS spectra of red giant branch stars in M31's dwarf spheroidal galaxies. We validate our method using high quality data that cover a range of S/N regimes: MW globular clusters (high S/N), M31 dSphs with deep observations (lower S/N), and M31 dSphs with shallow observations (lowest S/N). 

For both our validation coadds and science coadds, we include only stars that are known members of a given object (either GC or dSph), selected using their radial velocities, position on the CMD, and spectral features. We show that we can accurately reproduce the weighted average \feh\ and \alphafe\ of coadded spectra, where similar stars grouped by either photometric temperature (\teff) or metallicity (\feh). We find no significant differences in the results with either binning method, and we choose to present the measurements binned by photometric \feh. Using validation coadds, we characterize the systematic uncertainties of our coadded measurements, and we find that the systematic uncertainty limit that we adopt from \citet{kirby10} is sufficient.

We then coadd all of the member stars in our M31 dSph slitmasks that did not pass the quality criteria, i.e., they have large uncertainties, they hit the edge of the spectral synthesis grid during the continuum refinement step, or in the continuum refinement did not converge. We compare the science coadds to both the 1D (\feh) and 2D (\alphafe--\feh) abundance distributions from measurements of individual stars, and find that they agree. In \alphafe--\feh\ space, we find that our science coadds follow the same trend of decreasing \alphafe\ with increasing \feh, allowing us to fit slopes to these trends. 


For four of the dSphs in our sample (And XVIII, XV, XIV, IX), we present the first \alphafe\ measurements available. These galaxies, which are less massive and have fewer member stars than the rest of our sample, still show the same general trend of decreasing average metallicity as a function of stellar mass. In addition, there is some indication that these dSphs exhibit the same declining \alphafe-\feh\ trends that we find in the more massive dSphs. This trend indicates an extended star formation history for these less massive dSphs, consistent with what has been found via photometric studies.

Finally, we use our validation sample as well as our coadds to measure the average metallicity (\feh) of a dwarf galaxy as a function of mass. For four galaxies in our sample (And IX, XIV, XV, XVIII), these measurements are updated from the values presented in \citet{kirby13}, and agree within $1\sigma$. We also compare our values measured from combined coadded and individual measurements with those presented in \citet{kirby20}, measured using only individual measurements. Again, we find them to agree within $1\sigma$ ($\sim 0.2$ dex). We conclude that we can robustly determine average \feh\ for the dSphs in our sample, using combined coadded and individual measurements.

In future work, we plan to apply this coaddition method to stars across a number of M31 halo fields. These fields are too sparsely populated by member stars with high enough S/N to measure their abundances individually, but with our coaddition method, we will be able to measure average spectroscopic metallicities across the M31 halo. This will allow us to compare the metallicity, and therefore formation history, of the smooth halo compared to its rich substructure.

\acknowledgments
We thank the anonymous referee for their thorough comments and suggestions, which have greatly improved the quality of the manuscript. 

This material is based upon work supported by the National Science
Foundation under Grant Nos.\ AST-1614569 (JW, KMG) and AST-1614081 (ENK, IE).  ENK
gratefully acknowledges support from a Cottrell Scholar award
administered by the Research Corporation for Science Advancement as
well as funding from generous donors to the California Institute of
Technology.  IE acknowledges support from a National Science
Foundation (NSF) Graduate Research Fellowship under Grant
No.\ DGE-1745301. RLB and SRM thank NSF grants AST-0307842, AST-0607726, AST-1009882, and AST-1413269. 

The data presented herein were obtained at the W. M. Keck Observatory, which is operated as a scientific partnership among the California Institute of Technology, the University of California and the National Aeronautics and Space Administration. The Observatory was made possible by the generous financial support of the W. M. Keck Foundation.

The authors wish to recognize and acknowledge the very significant cultural role and reverence that the summit of Mauna Kea has always had within the indigenous Hawaiian community.  We are most fortunate to have the opportunity to conduct observations from this mountain.

This research made use of Astropy, a community-developed core Python package for Astronomy \citep{astropy2013,astropy2018}\footnote{http://www.astropy.org}. 
 
\facilities{Keck:II (DEIMOS)}
\software{Astropy \citep{astropy2013},
Matplotlib \citep{matplotlib}, numpy \citep{numpy}, scipy \citep{scipy}, emcee \citep{foreman-mackey13}}

\bibliography{mybib}{}

\begin{thebibliography}{}
\expandafter\ifx\csname natexlab\endcsname\relax\def\natexlab#1{#1}\fi
\providecommand{\url}[1]{\href{#1}{#1}}
\providecommand{\dodoi}[1]{doi:~\href{http://doi.org/#1}{\nolinkurl{#1}}}
\providecommand{\doeprint}[1]{\href{http://ascl.net/#1}{\nolinkurl{http://ascl.net/#1}}}
\providecommand{\doarXiv}[1]{\href{https://arxiv.org/abs/#1}{\nolinkurl{https://arxiv.org/abs/#1}}}

\bibitem[{{Adelman-McCarthy} {et~al.}(2006){Adelman-McCarthy}, {Ag{\"u}eros},
  {Allam}, {Anderson}, {Anderson}, {Annis}, {Bahcall}, {Baldry}, {Barentine},
  {Berlind}, {Bernardi}, {Blanton}, {Boroski}, {Brewington}, {Brinchmann},
  {Brinkmann}, {Brunner}, {Budav{\'a}ri}, {Carey}, {Carr}, {Castander},
  {Connolly}, {Csabai}, {Czarapata}, {Dalcanton}, {Doi}, {Dong}, {Eisenstein},
  {Evans}, {Fan}, {Finkbeiner}, {Friedman}, {Frieman}, {Fukugita}, {Gillespie},
  {Glazebrook}, {Gray}, {Grebel}, {Gunn}, {Gurbani}, {de Haas}, {Hall},
  {Harris}, {Harvanek}, {Hawley}, {Hayes}, {Hendry}, {Hennessy}, {Hindsley},
  {Hirata}, {Hogan}, {Hogg}, {Holmgren}, {Holtzman}, {Ichikawa}, {Ivezi{\'c}},
  {Jester}, {Johnston}, {Jorgensen}, {Juri{\'c}}, {Kent}, {Kleinman}, {Knapp},
  {Kniazev}, {Kron}, {Krzesinski}, {Kuropatkin}, {Lamb}, {Lampeitl}, {Lee},
  {Leger}, {Lin}, {Long}, {Loveday}, {Lupton}, {Margon},
  {Mart{\'\i}nez-Delgado}, {Mand elbaum}, {Matsubara}, {McGehee}, {McKay},
  {Meiksin}, {Munn}, {Nakajima}, {Nash}, {Neilsen}, {Newberg}, {Newman},
  {Nichol}, {Nicinski}, {Nieto-Santisteban}, {Nitta}, {O'Mullane}, {Okamura},
  {Owen}, {Padmanabhan}, {Pauls}, {Peoples}, {Pier}, {Pope}, {Pourbaix},
  {Quinn}, {Richards}, {Richmond}, {Rockosi}, {Schlegel}, {Schneider},
  {Schroeder}, {Scranton}, {Seljak}, {Sheldon}, {Shimasaku}, {Smith},
  {Smol{\v{c}}i{\'c}}, {Snedden}, {Stoughton}, {Strauss}, {SubbaRao}, {Szalay},
  {Szapudi}, {Szkody}, {Tegmark}, {Thakar}, {Tucker}, {Uomoto}, {Vanden Berk},
  {Vandenberg}, {Vogeley}, {Voges}, {Vogt}, {Walkowicz}, {Weinberg}, {West},
  {White}, {Xu}, {Yanny}, {Yocum}, {York}, {Zehavi}, {Zibetti}, \&
  {Zucker}}]{adelman-mccarthy06}
{Adelman-McCarthy}, J.~K., {Ag{\"u}eros}, M.~A., {Allam}, S.~S., {et~al.} 2006,
  \apjs, 162, 38, \dodoi{10.1086/497917}

\bibitem[{{Arnett}(1996)}]{arnett96}
{Arnett}, D. 1996, {Supernovae and nucleosynthesis. an investigation of the
  history of matter, from the Big Bang to the present}

\bibitem[{{Astropy Collaboration} {et~al.}(2013){Astropy Collaboration},
  {Robitaille}, {Tollerud}, {Greenfield}, {Droettboom}, {Bray}, {Aldcroft},
  {Davis}, {Ginsburg}, {Price-Whelan}, {Kerzendorf}, {Conley}, {Crighton},
  {Barbary}, {Muna}, {Ferguson}, {Grollier}, {Parikh}, {Nair}, {Unther},
  {Deil}, {Woillez}, {Conseil}, {Kramer}, {Turner}, {Singer}, {Fox}, {Weaver},
  {Zabalza}, {Edwards}, {Azalee Bostroem}, {Burke}, {Casey}, {Crawford},
  {Dencheva}, {Ely}, {Jenness}, {Labrie}, {Lim}, {Pierfederici}, {Pontzen},
  {Ptak}, {Refsdal}, {Servillat}, \& {Streicher}}]{astropy2013}
{Astropy Collaboration}, {Robitaille}, T.~P., {Tollerud}, E.~J., {et~al.} 2013,
  \aap, 558, A33, \dodoi{10.1051/0004-6361/201322068}

\bibitem[{{Astropy Collaboration} {et~al.}(2018){Astropy Collaboration},
  {Price-Whelan}, {Sip{\H{o}}cz}, {G{\"u}nther}, {Lim}, {Crawford}, {Conseil},
  {Shupe}, {Craig}, {Dencheva}, {Ginsburg}, {Vand erPlas}, {Bradley},
  {P{\'e}rez-Su{\'a}rez}, {de Val-Borro}, {Aldcroft}, {Cruz}, {Robitaille},
  {Tollerud}, {Ardelean}, {Babej}, {Bach}, {Bachetti}, {Bakanov}, {Bamford},
  {Barentsen}, {Barmby}, {Baumbach}, {Berry}, {Biscani}, {Boquien}, {Bostroem},
  {Bouma}, {Brammer}, {Bray}, {Breytenbach}, {Buddelmeijer}, {Burke},
  {Calderone}, {Cano Rodr{\'\i}guez}, {Cara}, {Cardoso}, {Cheedella}, {Copin},
  {Corrales}, {Crichton}, {D'Avella}, {Deil}, {Depagne}, {Dietrich}, {Donath},
  {Droettboom}, {Earl}, {Erben}, {Fabbro}, {Ferreira}, {Finethy}, {Fox},
  {Garrison}, {Gibbons}, {Goldstein}, {Gommers}, {Greco}, {Greenfield},
  {Groener}, {Grollier}, {Hagen}, {Hirst}, {Homeier}, {Horton}, {Hosseinzadeh},
  {Hu}, {Hunkeler}, {Ivezi{\'c}}, {Jain}, {Jenness}, {Kanarek}, {Kendrew},
  {Kern}, {Kerzendorf}, {Khvalko}, {King}, {Kirkby}, {Kulkarni}, {Kumar},
  {Lee}, {Lenz}, {Littlefair}, {Ma}, {Macleod}, {Mastropietro}, {McCully},
  {Montagnac}, {Morris}, {Mueller}, {Mumford}, {Muna}, {Murphy}, {Nelson},
  {Nguyen}, {Ninan}, {N{\"o}the}, {Ogaz}, {Oh}, {Parejko}, {Parley}, {Pascual},
  {Patil}, {Patil}, {Plunkett}, {Prochaska}, {Rastogi}, {Reddy Janga},
  {Sabater}, {Sakurikar}, {Seifert}, {Sherbert}, {Sherwood-Taylor}, {Shih},
  {Sick}, {Silbiger}, {Singanamalla}, {Singer}, {Sladen}, {Sooley},
  {Sornarajah}, {Streicher}, {Teuben}, {Thomas}, {Tremblay}, {Turner},
  {Terr{\'o}n}, {van Kerkwijk}, {de la Vega}, {Watkins}, {Weaver}, {Whitmore},
  {Woillez}, {Zabalza}, \& {Astropy Contributors}}]{astropy2018}
{Astropy Collaboration}, {Price-Whelan}, A.~M., {Sip{\H{o}}cz}, B.~M., {et~al.}
  2018, \aj, 156, 123, \dodoi{10.3847/1538-3881/aabc4f}

\bibitem[{{Beaton}(2014)}]{beaton14}
{Beaton}, R.~L. 2014, PhD thesis, University of Virginia

\bibitem[{{Boeche} {et~al.}(2013){Boeche}, {Siebert}, {Piffl}, {Just},
  {Steinmetz}, {Sharma}, {Kordopatis}, {Gilmore}, {Chiappini}, {Williams},
  {Grebel}, {Bland-Hawthorn}, {Gibson}, {Munari}, {Siviero}, {Bienaym{\'e}},
  {Navarro}, {Parker}, {Reid}, {Seabroke}, {Watson}, {Wyse}, \&
  {Zwitter}}]{Boeche13b}
{Boeche}, C., {Siebert}, A., {Piffl}, T., {et~al.} 2013, \aap, 559, A59,
  \dodoi{10.1051/0004-6361/201322085}

\bibitem[{{Bullock} \& {Johnston}(2005)}]{Bullock05}
{Bullock}, J.~S., \& {Johnston}, K.~V. 2005, \apj, 635, 931,
  \dodoi{10.1086/497422}

\bibitem[{{Carollo} {et~al.}(2007){Carollo}, {Beers}, {Lee}, {Chiba}, {Norris},
  {Wilhelm}, {Sivarani}, {Marsteller}, {Munn}, {Bailer-Jones}, {Fiorentin}, \&
  {York}}]{Carollo07}
{Carollo}, D., {Beers}, T.~C., {Lee}, Y.~S., {et~al.} 2007, \nat, 450, 1020,
  \dodoi{10.1038/nature06460}

\bibitem[{{Castelli} {et~al.}(1997){Castelli}, {Gratton}, \&
  {Kurucz}}]{castelli97}
{Castelli}, F., {Gratton}, R.~G., \& {Kurucz}, R.~L. 1997, \aap, 318, 841

\bibitem[{{Collins} {et~al.}(2011){Collins}, {Chapman}, {Rich}, {Irwin},
  {Pe{\~n}arrubia}, {Ibata}, {Arimoto}, {Brooks}, {Ferguson}, {Lewis},
  {McConnachie}, \& {Venn}}]{collins2011}
{Collins}, M.~L.~M., {Chapman}, S.~C., {Rich}, R.~M., {et~al.} 2011, \mnras,
  417, 1170, \dodoi{10.1111/j.1365-2966.2011.19342.x}

\bibitem[{{Collins} {et~al.}(2013){Collins}, {Chapman}, {Rich}, {Ibata},
  {Martin}, {Irwin}, {Bate}, {Lewis}, {Pe{\~n}arrubia}, {Arimoto}, {Casey},
  {Ferguson}, {Koch}, {McConnachie}, \& {Tanvir}}]{collins2013}
{Collins}, M. L.~M., {Chapman}, S.~C., {Rich}, R.~M., {et~al.} 2013, \apj, 768,
  172, \dodoi{10.1088/0004-637X/768/2/172}

\bibitem[{{Cooper} {et~al.}(2010){Cooper}, {Cole}, {Frenk}, {White}, {Helly},
  {Benson}, {De Lucia}, {Helmi}, {Jenkins}, {Navarro}, {Springel}, \&
  {Wang}}]{cooper10}
{Cooper}, A.~P., {Cole}, S., {Frenk}, C.~S., {et~al.} 2010, \mnras, 406, 744,
  \dodoi{10.1111/j.1365-2966.2010.16740.x}

\bibitem[{{Cooper} {et~al.}(2012){Cooper}, {Newman}, {Davis}, {Finkbeiner}, \&
  {Gerke}}]{cooper12}
{Cooper}, M.~C., {Newman}, J.~A., {Davis}, M., {Finkbeiner}, D.~P., \& {Gerke},
  B.~F. 2012, {spec2d: DEEP2 DEIMOS Spectral Pipeline}, Astrophysics Source
  Code Library.
\newblock \doeprint{1203.003}

\bibitem[{{Da Costa} {et~al.}(2002){Da Costa}, {Armandroff}, \&
  {Caldwell}}]{dacosta02}
{Da Costa}, G.~S., {Armandroff}, T.~E., \& {Caldwell}, N. 2002, \aj, 124, 332,
  \dodoi{10.1086/340965}

\bibitem[{{da Costa} {et~al.}(2000){da Costa}, {Bernardi}, {Alonso}, {Wegner},
  {Willmer}, {Pellegrini}, {Rit{\'e}}, \& {Maia}}]{dacosta00}
{da Costa}, L.~N., {Bernardi}, M., {Alonso}, M.~V., {et~al.} 2000, \aj, 120,
  95, \dodoi{10.1086/301449}

\bibitem[{{Demarque} {et~al.}(2004){Demarque}, {Woo}, {Kim}, \&
  {Yi}}]{demarque04}
{Demarque}, P., {Woo}, J.-H., {Kim}, Y.-C., \& {Yi}, S.~K. 2004, \apjs, 155,
  667, \dodoi{10.1086/424966}

\bibitem[{{Dorman} {et~al.}(2012){Dorman}, {Guhathakurta}, {Fardal}, {Lang},
  {Geha}, {Howley}, {Kalirai}, {Bullock}, {Cuillandre}, {Dalcanton}, {Gilbert},
  {Seth}, {Tollerud}, {Williams}, \& {Yniguez}}]{dorman12}
{Dorman}, C.~E., {Guhathakurta}, P., {Fardal}, M.~A., {et~al.} 2012, \apj, 752,
  147, \dodoi{10.1088/0004-637X/752/2/147}

\bibitem[{{Dorman} {et~al.}(2015){Dorman}, {Guhathakurta}, {Seth}, {Weisz},
  {Bell}, {Dalcanton}, {Gilbert}, {Hamren}, {Lewis}, {Skillman}, {Toloba}, \&
  {Williams}}]{dorman15}
{Dorman}, C.~E., {Guhathakurta}, P., {Seth}, A.~C., {et~al.} 2015, \apj, 803,
  24, \dodoi{10.1088/0004-637X/803/1/24}

\bibitem[{{Escala} {et~al.}(2019{\natexlab{a}}){Escala}, {Gilbert}, {Kirby},
  {Wojno}, {Cunningham}, \& {Guhathakurta}}]{escala19b}
{Escala}, I., {Gilbert}, K.~M., {Kirby}, E.~N., {et~al.} 2019{\natexlab{a}},
  arXiv e-prints, arXiv:1909.00006.
\newblock \doarXiv{1909.00006}

\bibitem[{{Escala} {et~al.}(2019{\natexlab{b}}){Escala}, {Kirby}, {Gilbert},
  {Cunningham}, \& {Wojno}}]{escala18}
{Escala}, I., {Kirby}, E.~N., {Gilbert}, K.~M., {Cunningham}, E.~C., \&
  {Wojno}, J. 2019{\natexlab{b}}, \apj, 878, 42,
  \dodoi{10.3847/1538-4357/ab1eac}

\bibitem[{{Faber} {et~al.}(2003){Faber}, {Phillips}, {Kibrick}, {Alcott},
  {Allen}, {Burrous}, {Cantrall}, {Clarke}, {Coil}, {Cowley}, {Davis}, {Deich},
  {Dietsch}, {Gilmore}, {Harper}, {Hilyard}, {Lewis}, {McVeigh}, {Newman},
  {Osborne}, {Schiavon}, {Stover}, {Tucker}, {Wallace}, {Wei}, {Wirth}, \&
  {Wright}}]{faber03}
{Faber}, S.~M., {Phillips}, A.~C., {Kibrick}, R.~I., {et~al.} 2003, in
  \procspie, Vol. 4841, Instrument Design and Performance for Optical/Infrared
  Ground-based Telescopes, ed. M.~{Iye} \& A.~F.~M. {Moorwood}, 1657--1669,
  \dodoi{10.1117/12.460346}

\bibitem[{{Font} {et~al.}(2011){Font}, {McCarthy}, {Crain}, {Theuns}, {Schaye},
  {Wiersma}, \& {Dalla Vecchia}}]{font11}
{Font}, A.~S., {McCarthy}, I.~G., {Crain}, R.~A., {et~al.} 2011, \mnras, 416,
  2802, \dodoi{10.1111/j.1365-2966.2011.19227.x}

\bibitem[{{Foreman-Mackey} {et~al.}(2013){Foreman-Mackey}, {Hogg}, {Lang}, \&
  {Goodman}}]{foreman-mackey13}
{Foreman-Mackey}, D., {Hogg}, D.~W., {Lang}, D., \& {Goodman}, J. 2013, \pasp,
  125, 306, \dodoi{10.1086/670067}

\bibitem[{{Gilbert} {et~al.}(2019){Gilbert}, {Kirby}, {Escala}, {Wojno},
  {Kalirai}, \& {Guhathakurta}}]{gilbert19a}
{Gilbert}, K.~M., {Kirby}, E.~N., {Escala}, I., {et~al.} 2019, \apj, 883, 128,
  \dodoi{10.3847/1538-4357/ab3807}

\bibitem[{{Gilbert} {et~al.}(2006){Gilbert}, {Guhathakurta}, {Kalirai}, {Rich},
  {Majewski}, {Ostheimer}, {Reitzel}, {Cenarro}, {Cooper}, {Luine}, \&
  {Patterson}}]{gilbert06}
{Gilbert}, K.~M., {Guhathakurta}, P., {Kalirai}, J.~S., {et~al.} 2006, \apj,
  652, 1188, \dodoi{10.1086/508643}

\bibitem[{{Gilbert} {et~al.}(2012){Gilbert}, {Guhathakurta}, {Beaton},
  {Bullock}, {Geha}, {Kalirai}, {Kirby}, {Majewski}, {Ostheimer}, {Patterson},
  {Tollerud}, {Tanaka}, \& {Chiba}}]{gilbert12}
{Gilbert}, K.~M., {Guhathakurta}, P., {Beaton}, R.~L., {et~al.} 2012, \apj,
  760, 76, \dodoi{10.1088/0004-637X/760/1/76}

\bibitem[{{Gilbert} {et~al.}(2014){Gilbert}, {Kalirai}, {Guhathakurta},
  {Beaton}, {Geha}, {Kirby}, {Majewski}, {Patterson}, {Tollerud}, {Bullock},
  {Tanaka}, \& {Chiba}}]{gilbert14}
{Gilbert}, K.~M., {Kalirai}, J.~S., {Guhathakurta}, P., {et~al.} 2014, \apj,
  796, 76, \dodoi{10.1088/0004-637X/796/2/76}

\bibitem[{{Gilbert} {et~al.}(2018){Gilbert}, {Tollerud}, {Beaton},
  {Guhathakurta}, {Bullock}, {Chiba}, {Kalirai}, {Kirby}, {Majewski}, \&
  {Tanaka}}]{gilbert18}
{Gilbert}, K.~M., {Tollerud}, E., {Beaton}, R.~L., {et~al.} 2018, \apj, 852,
  128, \dodoi{10.3847/1538-4357/aa9f26}

\bibitem[{{Gilmore} \& {Wyse}(1991)}]{gilmore91}
{Gilmore}, G., \& {Wyse}, R. F.~G. 1991, \apjl, 367, L55,
  \dodoi{10.1086/185930}

\bibitem[{{Gilmore} \& {Wyse}(1998)}]{gilmore98}
---. 1998, \aj, 116, 748, \dodoi{10.1086/300459}

\bibitem[{{Gilmore} {et~al.}(2002){Gilmore}, {Wyse}, \& {Norris}}]{gilmore02}
{Gilmore}, G., {Wyse}, R.~F.~G., \& {Norris}, J.~E. 2002, \apjl, 574, L39,
  \dodoi{10.1086/342363}

\bibitem[{{Girardi}(2016)}]{Girardi16}
{Girardi}, L. 2016, \araa, 54, 95, \dodoi{10.1146/annurev-astro-081915-023354}

\bibitem[{{Grebel} \& {Guhathakurta}(1999)}]{grebel99}
{Grebel}, E.~K., \& {Guhathakurta}, P. 1999, \apjl, 511, L101,
  \dodoi{10.1086/311852}

\bibitem[{{Guhathakurta} {et~al.}(2005){Guhathakurta}, {Ostheimer}, {Gilbert},
  {Rich}, {Majewski}, {Kalirai}, {Reitzel}, \& {Patterson}}]{guhathakurta05}
{Guhathakurta}, P., {Ostheimer}, J.~C., {Gilbert}, K.~M., {et~al.} 2005, arXiv
  e-prints, astro.
\newblock \doarXiv{astro-ph/0502366}

\bibitem[{{Guhathakurta} {et~al.}(2006){Guhathakurta}, {Rich}, {Reitzel},
  {Cooper}, {Gilbert}, {Majewski}, {Ostheimer}, {Geha}, {Johnston}, \&
  {Patterson}}]{guhathakurta06}
{Guhathakurta}, P., {Rich}, R.~M., {Reitzel}, D.~B., {et~al.} 2006, \aj, 131,
  2497, \dodoi{10.1086/499562}

\bibitem[{{Hayden} {et~al.}(2015){Hayden}, {Bovy}, {Holtzman}, {Nidever},
  {Bird}, {Weinberg}, {Andrews}, {Majewski}, {Allende Prieto}, {Anders},
  {Beers}, {Bizyaev}, {Chiappini}, {Cunha}, {Frinchaboy},
  {Garc{\'{\i}}a-Her{\'n}andez}, {Garc{\'{\i}}a P{\'e}rez}, {Girardi},
  {Harding}, {Hearty}, {Johnson}, {M{\'e}sz{\'a}ros}, {Minchev}, {O'Connell},
  {Pan}, {Robin}, {Schiavon}, {Schneider}, {Schultheis}, {Shetrone},
  {Skrutskie}, {Steinmetz}, {Smith}, {Wilson}, {Zamora}, \&
  {Zasowski}}]{Hayden15}
{Hayden}, M.~R., {Bovy}, J., {Holtzman}, J.~A., {et~al.} 2015, \apj, 808, 132,
  \dodoi{10.1088/0004-637X/808/2/132}

\bibitem[{{Helmi} {et~al.}(2017){Helmi}, {Veljanoski}, {Breddels}, {Tian}, \&
  {Sales}}]{Helmi17}
{Helmi}, A., {Veljanoski}, J., {Breddels}, M.~A., {Tian}, H., \& {Sales}, L.~V.
  2017, \aap, 598, A58, \dodoi{10.1051/0004-6361/201629990}

\bibitem[{{Ho} {et~al.}(2012){Ho}, {Geha}, {Munoz}, {Guhathakurta}, {Kalirai},
  {Gilbert}, {Tollerud}, {Bullock}, {Beaton}, \& {Majewski}}]{ho12}
{Ho}, N., {Geha}, M., {Munoz}, R.~R., {et~al.} 2012, \apj, 758, 124,
  \dodoi{10.1088/0004-637X/758/2/124}

\bibitem[{{Hogg} {et~al.}(2010){Hogg}, {Bovy}, \& {Lang}}]{hogg10}
{Hogg}, D.~W., {Bovy}, J., \& {Lang}, D. 2010, arXiv e-prints, arXiv:1008.4686.
\newblock \doarXiv{1008.4686}

\bibitem[{{Hunter}(2007)}]{matplotlib}
{Hunter}, J.~D. 2007, Computing in Science and Engineering, 9, 90,
  \dodoi{10.1109/MCSE.2007.55}

\bibitem[{{Ibata} {et~al.}(2001){Ibata}, {Irwin}, {Lewis}, {Ferguson}, \&
  {Tanvir}}]{Ibata01}
{Ibata}, R., {Irwin}, M., {Lewis}, G., {Ferguson}, A. M.~N., \& {Tanvir}, N.
  2001, Nature, 412, 49.
\newblock \doarXiv{astro-ph/0107090}

\bibitem[{{Ibata} {et~al.}(1994){Ibata}, {Gilmore}, \& {Irwin}}]{Ibata94}
{Ibata}, R.~A., {Gilmore}, G., \& {Irwin}, M.~J. 1994, \nat, 370, 194,
  \dodoi{10.1038/370194a0}

\bibitem[{{Ibata} {et~al.}(2013){Ibata}, {Lewis}, {Conn}, {Irwin},
  {McConnachie}, {Chapman}, {Collins}, {Fardal}, {Ferguson}, {Ibata}, {Mackey},
  {Martin}, {Navarro}, {Rich}, {Valls-Gabaud}, \& {Widrow}}]{Ibata13}
{Ibata}, R.~A., {Lewis}, G.~F., {Conn}, A.~R., {et~al.} 2013, \nat, 493, 62,
  \dodoi{10.1038/nature11717}

\bibitem[{{Ishigaki} {et~al.}(2012){Ishigaki}, {Chiba}, \& {Aoki}}]{Ishigaki12}
{Ishigaki}, M.~N., {Chiba}, M., \& {Aoki}, W. 2012, \apj, 753, 64,
  \dodoi{10.1088/0004-637X/753/1/64}

\bibitem[{{Kalirai} {et~al.}(2006){Kalirai}, {Gilbert}, {Guhathakurta},
  {Majewski}, {Ostheimer}, {Rich}, {Cooper}, {Reitzel}, \&
  {Patterson}}]{kalirai06}
{Kalirai}, J.~S., {Gilbert}, K.~M., {Guhathakurta}, P., {et~al.} 2006, \apj,
  648, 389, \dodoi{10.1086/505697}

\bibitem[{{Kalirai} {et~al.}(2009){Kalirai}, {Zucker}, {Guhathakurta}, {Geha},
  {Kniazev}, {Mart{\'\i}nez-Delgado}, {Bell}, {Grebel}, \&
  {Gilbert}}]{kalirai09}
{Kalirai}, J.~S., {Zucker}, D.~B., {Guhathakurta}, P., {et~al.} 2009, \apj,
  705, 1043, \dodoi{10.1088/0004-637X/705/1/1043}

\bibitem[{{Kalirai} {et~al.}(2010){Kalirai}, {Beaton}, {Geha}, {Gilbert},
  {Guhathakurta}, {Kirby}, {Majewski}, {Ostheimer}, {Patterson}, \&
  {Wolf}}]{kalirai10}
{Kalirai}, J.~S., {Beaton}, R.~L., {Geha}, M.~C., {et~al.} 2010, \apj, 711,
  671, \dodoi{10.1088/0004-637X/711/2/671}

\bibitem[{{Kirby}(2011)}]{kirby11}
{Kirby}, E.~N. 2011, \pasp, 123, 531, \dodoi{10.1086/660019}

\bibitem[{{Kirby} {et~al.}(2013){Kirby}, {Cohen}, {Guhathakurta}, {Cheng},
  {Bullock}, \& {Gallazzi}}]{kirby13}
{Kirby}, E.~N., {Cohen}, J.~G., {Guhathakurta}, P., {et~al.} 2013, \apj, 779,
  102, \dodoi{10.1088/0004-637X/779/2/102}

\bibitem[{{Kirby} {et~al.}(2020){Kirby}, {Gilbert}, {Escala}, {Wojno},
  {Guhathakurta}, {Majewski}, \& {Beaton}}]{kirby20}
{Kirby}, E.~N., {Gilbert}, K.~M., {Escala}, I., {et~al.} 2020, \aj, 159, 46,
  \dodoi{10.3847/1538-3881/ab5f0f}

\bibitem[{{Kirby} {et~al.}(2009){Kirby}, {Guhathakurta}, {Bolte}, {Sneden}, \&
  {Geha}}]{kirby09}
{Kirby}, E.~N., {Guhathakurta}, P., {Bolte}, M., {Sneden}, C., \& {Geha}, M.~C.
  2009, \apj, 705, 328, \dodoi{10.1088/0004-637X/705/1/328}

\bibitem[{{Kirby} {et~al.}(2008){Kirby}, {Guhathakurta}, \&
  {Sneden}}]{kirby08a}
{Kirby}, E.~N., {Guhathakurta}, P., \& {Sneden}, C. 2008, \apj, 682, 1217,
  \dodoi{10.1086/589627}

\bibitem[{{Kirby} {et~al.}(2016){Kirby}, {Guhathakurta}, {Zhang}, {Hong},
  {Guo}, {Guo}, {Cohen}, \& {Cunha}}]{kirby16}
{Kirby}, E.~N., {Guhathakurta}, P., {Zhang}, A.~J., {et~al.} 2016, \apj, 819,
  135, \dodoi{10.3847/0004-637X/819/2/135}

\bibitem[{{Kirby} {et~al.}(2015){Kirby}, {Simon}, \& {Cohen}}]{kirby15}
{Kirby}, E.~N., {Simon}, J.~D., \& {Cohen}, J.~G. 2015, \apj, 810, 56,
  \dodoi{10.1088/0004-637X/810/1/56}

\bibitem[{{Kirby} {et~al.}(2010){Kirby}, {Guhathakurta}, {Simon}, {Geha},
  {Rockosi}, {Sneden}, {Cohen}, {Sohn}, {Majewski}, \& {Siegel}}]{kirby10}
{Kirby}, E.~N., {Guhathakurta}, P., {Simon}, J.~D., {et~al.} 2010, \apjs, 191,
  352, \dodoi{10.1088/0067-0049/191/2/352}

\bibitem[{{Kirby} {et~al.}(2019){Kirby}, {Xie}, {Guo}, {de los Reyes},
  {Bergemann}, {Kovalev}, {Shen}, {Piro}, \& {McWilliam}}]{kirby19ni}
{Kirby}, E.~N., {Xie}, J.~L., {Guo}, R., {et~al.} 2019, \apj, 881, 45,
  \dodoi{10.3847/1538-4357/ab2c02}

\bibitem[{{Kordopatis} {et~al.}(2015){Kordopatis}, {Binney}, {Gilmore}, {Wyse},
  {Belokurov}, {McMillan}, {Hatfield}, {Grebel}, {Steinmetz}, {Navarro},
  {Seabroke}, {Minchev}, {Chiappini}, {Bienaym{\'e}}, {Bland-Hawthorn},
  {Freeman}, {Gibson}, {Helmi}, {Munari}, {Parker}, {Reid}, {Siebert},
  {Siviero}, \& {Zwitter}}]{Kordopatis15_rich}
{Kordopatis}, G., {Binney}, J., {Gilmore}, G., {et~al.} 2015, \mnras, 447,
  3526, \dodoi{10.1093/mnras/stu2726}

\bibitem[{{Kupka} {et~al.}(1999){Kupka}, {Piskunov}, {Ryabchikova}, {Stempels},
  \& {Weiss}}]{kupka99}
{Kupka}, F., {Piskunov}, N., {Ryabchikova}, T.~A., {Stempels}, H.~C., \&
  {Weiss}, W.~W. 1999, \aaps, 138, 119, \dodoi{10.1051/aas:1999267}

\bibitem[{{Kurucz}(1992)}]{kurucz92}
{Kurucz}, R.~L. 1992, \rmxaa, 23

\bibitem[{{Kurucz}(1993)}]{kurucz93}
---. 1993, Physica Scripta Volume T, 47, 110,
  \dodoi{10.1088/0031-8949/1993/T47/017}

\bibitem[{{Letarte} {et~al.}(2010){Letarte}, {Hill}, {Tolstoy}, {Jablonka},
  {Shetrone}, {Venn}, {Spite}, {Irwin}, {Battaglia}, {Helmi}, {Primas},
  {Fran{\c{c}}ois}, {Kaufer}, {Szeifert}, {Arimoto}, \& {Sadakane}}]{letarte10}
{Letarte}, B., {Hill}, V., {Tolstoy}, E., {et~al.} 2010, \aap, 523, A17,
  \dodoi{10.1051/0004-6361/200913413}

\bibitem[{{Lynden-Bell}(1975)}]{lynden-bell75}
{Lynden-Bell}, D. 1975, Vistas in Astronomy, 19, 299,
  \dodoi{10.1016/0083-6656(75)90005-7}

\bibitem[{{Majewski} {et~al.}(2000){Majewski}, {Ostheimer}, {Kunkel}, \&
  {Patterson}}]{majewski00}
{Majewski}, S.~R., {Ostheimer}, J.~C., {Kunkel}, W.~E., \& {Patterson}, R.~J.
  2000, \aj, 120, 2550, \dodoi{10.1086/316836}

\bibitem[{{Martin} {et~al.}(2017){Martin}, {Weisz}, {Albers}, {Bernard},
  {Collins}, {Dolphin}, {Ferguson}, {Ibata}, {Laevens}, {Lewis}, {Mackey},
  {McConnachie}, {Rich}, \& {Skillman}}]{martin17}
{Martin}, N.~F., {Weisz}, D.~R., {Albers}, S.~M., {et~al.} 2017, \apj, 850, 16,
  \dodoi{10.3847/1538-4357/aa901a}

\bibitem[{{Matteucci}(2001)}]{Matteucci01}
{Matteucci}, F., ed. 2001, Astrophysics and Space Science Library, Vol. 253,
  {The chemical evolution of the Galaxy}, \dodoi{10.1007/978-94-010-0967-6}

\bibitem[{{McConnachie}(2012)}]{mcconnachie12}
{McConnachie}, A.~W. 2012, \aj, 144, 4, \dodoi{10.1088/0004-6256/144/1/4}

\bibitem[{{McConnachie} {et~al.}(2007){McConnachie}, {Arimoto}, \&
  {Irwin}}]{mcconnachie07}
{McConnachie}, A.~W., {Arimoto}, N., \& {Irwin}, M. 2007, \mnras, 379, 379,
  \dodoi{10.1111/j.1365-2966.2007.11969.x}

\bibitem[{{Newman} {et~al.}(2013){Newman}, {Cooper}, {Davis}, {Faber}, {Coil},
  {Guhathakurta}, {Koo}, {Phillips}, {Conroy}, {Dutton}, {Finkbeiner}, {Gerke},
  {Rosario}, {Weiner}, {Willmer}, {Yan}, {Harker}, {Kassin}, {Konidaris},
  {Lai}, {Madgwick}, {Noeske}, {Wirth}, {Connolly}, {Kaiser}, {Kirby},
  {Lemaux}, {Lin}, {Lotz}, {Luppino}, {Marinoni}, {Matthews}, {Metevier}, \&
  {Schiavon}}]{newman13}
{Newman}, J.~A., {Cooper}, M.~C., {Davis}, M., {et~al.} 2013, \apjs, 208, 5,
  \dodoi{10.1088/0067-0049/208/1/5}

\bibitem[{{Nomoto} {et~al.}(1997){Nomoto}, {Iwamoto}, {Nakasato}, {Thielemann},
  {Brachwitz}, {Tsujimoto}, {Kubo}, \& {Kishimoto}}]{nomoto97}
{Nomoto}, K., {Iwamoto}, K., {Nakasato}, N., {et~al.} 1997, Nuclear Physics A,
  621, 467, \dodoi{10.1016/S0375-9474(97)00291-1}

\bibitem[{{Ostheimer}(2003)}]{ostheimer03}
{Ostheimer}, James~Craig, J. 2003, PhD thesis, UNIVERSITY OF VIRGINIA

\bibitem[{{Pagel}(1997)}]{pagel97}
{Pagel}, B.~E.~J. 1997, {Nucleosynthesis and Chemical Evolution of Galaxies},
  392

\bibitem[{{Palma} {et~al.}(2003){Palma}, {Majewski}, {Siegel}, {Patterson},
  {Ostheimer}, \& {Link}}]{palma03}
{Palma}, C., {Majewski}, S.~R., {Siegel}, M.~H., {et~al.} 2003, \aj, 125, 1352,
  \dodoi{10.1086/367594}

\bibitem[{{Sbordone}(2005)}]{sbordone05}
{Sbordone}, L. 2005, Memorie della Societa Astronomica Italiana Supplementi, 8,
  61

\bibitem[{{Sbordone} {et~al.}(2004){Sbordone}, {Bonifacio}, {Castelli}, \&
  {Kurucz}}]{sbordone04}
{Sbordone}, L., {Bonifacio}, P., {Castelli}, F., \& {Kurucz}, R.~L. 2004,
  Memorie della Societa Astronomica Italiana Supplementi, 5, 93

\bibitem[{{Schmidt}(1963)}]{schmidt63}
{Schmidt}, M. 1963, \apj, 137, 758, \dodoi{10.1086/147553}

\bibitem[{{Simon} \& {Geha}(2007)}]{simon07}
{Simon}, J.~D., \& {Geha}, M. 2007, \apj, 670, 313, \dodoi{10.1086/521816}

\bibitem[{{Smecker-Hane} \& {Wyse}(1992)}]{smecker-hane92}
{Smecker-Hane}, T.~A., \& {Wyse}, R.~F.~G. 1992, \aj, 103, 1621,
  \dodoi{10.1086/116175}

\bibitem[{{Sneden}(1973)}]{sneden73}
{Sneden}, C.~A. 1973, PhD thesis, THE UNIVERSITY OF TEXAS AT AUSTIN.

\bibitem[{{Sohn} {et~al.}(2007){Sohn}, {Majewski}, {Mu{\~n}oz}, {Kunkel},
  {Johnston}, {Ostheimer}, {Guhathakurta}, {Patterson}, {Siegel}, \&
  {Cooper}}]{sohn07}
{Sohn}, S.~T., {Majewski}, S.~R., {Mu{\~n}oz}, R.~R., {et~al.} 2007, \apj, 663,
  960, \dodoi{10.1086/518302}

\bibitem[{{Talbot} \& {Arnett}(1971)}]{talbot71}
{Talbot}, Raymond~J., J., \& {Arnett}, W.~D. 1971, \apj, 170, 409,
  \dodoi{10.1086/151228}

\bibitem[{{Tinsley}(1979)}]{tinsley79}
{Tinsley}, B.~M. 1979, \apj, 229, 1046, \dodoi{10.1086/157039}

\bibitem[{{Tinsley}(1980)}]{tinsley80}
---. 1980, \fcp, 5, 287

\bibitem[{{Tollerud} {et~al.}(2013){Tollerud}, {Geha}, {Vargas}, \&
  {Bullock}}]{tollerud2013}
{Tollerud}, E.~J., {Geha}, M.~C., {Vargas}, L.~C., \& {Bullock}, J.~S. 2013,
  \apj, 768, 50, \dodoi{10.1088/0004-637X/768/1/50}

\bibitem[{{Tollerud} {et~al.}(2012){Tollerud}, {Beaton}, {Geha}, {Bullock},
  {Guhathakurta}, {Kalirai}, {Majewski}, {Kirby}, {Gilbert}, {Yniguez},
  {Patterson}, {Ostheimer}, {Cooke}, {Dorman}, {Choudhury}, \&
  {Cooper}}]{tollerud12}
{Tollerud}, E.~J., {Beaton}, R.~L., {Geha}, M.~C., {et~al.} 2012, \apj, 752,
  45, \dodoi{10.1088/0004-637X/752/1/45}

\bibitem[{{Tolstoy} {et~al.}(2009){Tolstoy}, {Hill}, \& {Tosi}}]{Tolstoy09}
{Tolstoy}, E., {Hill}, V., \& {Tosi}, M. 2009, \araa, 47, 371,
  \dodoi{10.1146/annurev-astro-082708-101650}

\bibitem[{{Tully} {et~al.}(2013){Tully}, {Courtois}, {Dolphin}, {Fisher},
  {H{\'e}raudeau}, {Jacobs}, {Karachentsev}, {Makarov}, {Makarova},
  {Mitronova}, {Rizzi}, {Shaya}, {Sorce}, \& {Wu}}]{Tully13}
{Tully}, R.~B., {Courtois}, H.~M., {Dolphin}, A.~E., {et~al.} 2013, The
  Astronomical Journal, 146, 86, \dodoi{10.1088/0004-6256/146/4/86}

\bibitem[{{van der Walt} {et~al.}(2011){van der Walt}, {Colbert}, \&
  {Varoquaux}}]{numpy}
{van der Walt}, S., {Colbert}, S.~C., \& {Varoquaux}, G. 2011, Computing in
  Science and Engineering, 13, 22, \dodoi{10.1109/MCSE.2011.37}

\bibitem[{{VandenBerg} {et~al.}(2006){VandenBerg}, {Bergbusch}, \&
  {Dowler}}]{vandenberg06}
{VandenBerg}, D.~A., {Bergbusch}, P.~A., \& {Dowler}, P.~D. 2006, \apjs, 162,
  375, \dodoi{10.1086/498451}

\bibitem[{{Vargas} {et~al.}(2013){Vargas}, {Geha}, {Kirby}, \&
  {Simon}}]{vargas13}
{Vargas}, L.~C., {Geha}, M., {Kirby}, E.~N., \& {Simon}, J.~D. 2013, \apj, 767,
  134, \dodoi{10.1088/0004-637X/767/2/134}

\bibitem[{{Vargas} {et~al.}(2014{\natexlab{a}}){Vargas}, {Geha}, \&
  {Tollerud}}]{vargas14a}
{Vargas}, L.~C., {Geha}, M.~C., \& {Tollerud}, E.~J. 2014{\natexlab{a}}, \apj,
  790, 73, \dodoi{10.1088/0004-637X/790/1/73}

\bibitem[{{Vargas} {et~al.}(2014{\natexlab{b}}){Vargas}, {Gilbert}, {Geha},
  {Tollerud}, {Kirby}, \& {Guhathakurta}}]{vargas14b}
{Vargas}, L.~C., {Gilbert}, K.~M., {Geha}, M., {et~al.} 2014{\natexlab{b}},
  \apjl, 797, L2, \dodoi{10.1088/2041-8205/797/1/L2}

\bibitem[{{Virtanen} {et~al.}(2019){Virtanen}, {Gommers}, {Oliphant},
  {Haberland}, {Reddy}, {Cournapeau}, {Burovski}, {Peterson}, {Weckesser},
  {Bright}, {van der Walt}, {Brett}, {Wilson}, {Jarrod Millman}, {Mayorov},
  {Nelson}, {Jones}, {Kern}, {Larson}, {Carey}, {Polat}, {Feng}, {Moore}, {Vand
  erPlas}, {Laxalde}, {Perktold}, {Cimrman}, {Henriksen}, {Quintero}, {Harris},
  {Archibald}, {Ribeiro}, {Pedregosa}, {van Mulbregt}, \&
  {Contributors}}]{scipy}
{Virtanen}, P., {Gommers}, R., {Oliphant}, T.~E., {et~al.} 2019, arXiv
  e-prints, arXiv:1907.10121.
\newblock \doarXiv{1907.10121}

\bibitem[{{Weisz} {et~al.}(2014){Weisz}, {Skillman}, {Hidalgo}, {Monelli},
  {Dolphin}, {McConnachie}, {Bernard}, {Gallart}, {Aparicio}, {Boylan-Kolchin},
  {Cassisi}, {Cole}, {Ferguson}, {Irwin}, {Martin}, {Mayer}, {McQuinn},
  {Navarro}, \& {Stetson}}]{weisz14a}
{Weisz}, D.~R., {Skillman}, E.~D., {Hidalgo}, S.~L., {et~al.} 2014, \apj, 789,
  24, \dodoi{10.1088/0004-637X/789/1/24}

\bibitem[{{Woo} {et~al.}(2008){Woo}, {Courteau}, \& {Dekel}}]{woo08}
{Woo}, J., {Courteau}, S., \& {Dekel}, A. 2008, \mnras, 390, 1453,
  \dodoi{10.1111/j.1365-2966.2008.13770.x}

\bibitem[{{Woosley} \& {Janka}(2005)}]{woosley05}
{Woosley}, S., \& {Janka}, T. 2005, Nature Physics, 1, 147,
  \dodoi{10.1038/nphys172}

\bibitem[{{Yang} {et~al.}(2013){Yang}, {Kirby}, {Guhathakurta}, {Peng}, \&
  {Cheng}}]{yang13}
{Yang}, L., {Kirby}, E.~N., {Guhathakurta}, P., {Peng}, E.~W., \& {Cheng}, L.
  2013, \apj, 768, 4, \dodoi{10.1088/0004-637X/768/1/4}

\bibitem[{{Zucker} {et~al.}(2007){Zucker}, {Kniazev}, {Mart{\'\i}nez-Delgado},
  {Bell}, {Rix}, {Grebel}, {Holtzman}, {Walterbos}, {Rockosi}, {York},
  {Barentine}, {Brewington}, {Brinkmann}, {Harvanek}, {Kleinman}, {Krzesinski},
  {Long}, {Neilsen}, {Nitta}, \& {Snedden}}]{zucker07}
{Zucker}, D.~B., {Kniazev}, A.~Y., {Mart{\'\i}nez-Delgado}, D., {et~al.} 2007,
  \apjl, 659, L21, \dodoi{10.1086/516748}

\end{thebibliography}
\bibliographystyle{aasjournal}

\appendix
\section{Comparison with Kirby et al. for abundance measurements from individual stars}
\label{sec:pipeline_comparison}

\begin{figure*}
    \centering
    \includegraphics[width=0.95\textwidth]{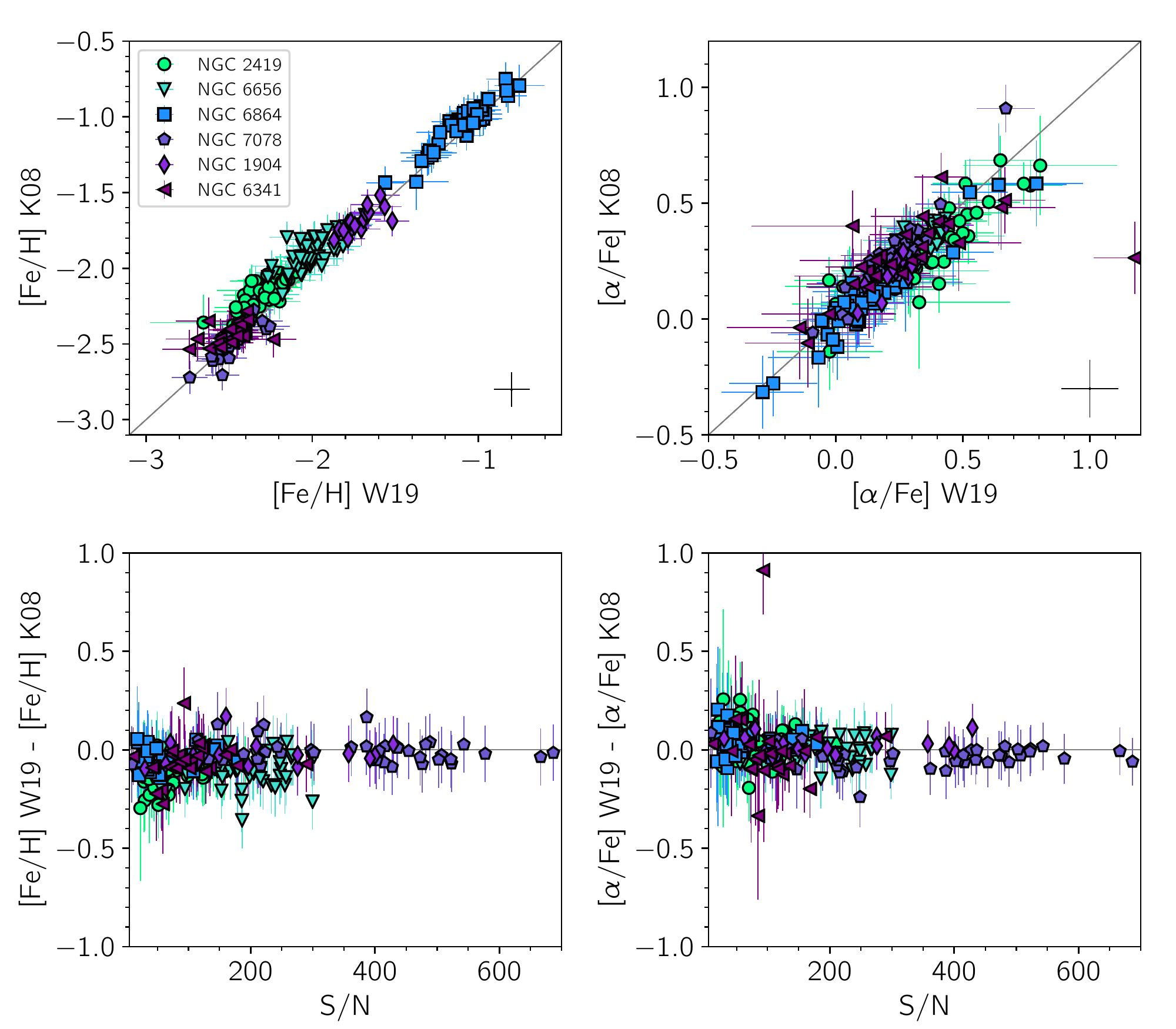}
    \caption{Top row: abundance comparison for all measurements for individual stars in our sample of MW GCs between the pipeline rewritten in Python \citep{escala18} used for this study (W19), and that used for previous studies, written in IDL \citep[K08,][]{kirby08a}. The left column shows \feh\ and the right column shows \alphafe\ for stars that pass the membership and quality criteria described in Section~\ref{sec:membership_quality_criteria}. Mean uncertainties are indicated in black for \feh\ and \alphafe\ in the left and right panels, respectively. Bottom row: the difference between the two pipelines as a function of S/N\@. We confirm that the two pipelines generally return similar results within the uncertainties, and the dispersion of the differences decreases as a function of increasing S/N.}
    \label{fig:ie_vs_enk_gcs}
\end{figure*}

\begin{figure*}
    \centering
 	\includegraphics[width=0.95\textwidth]{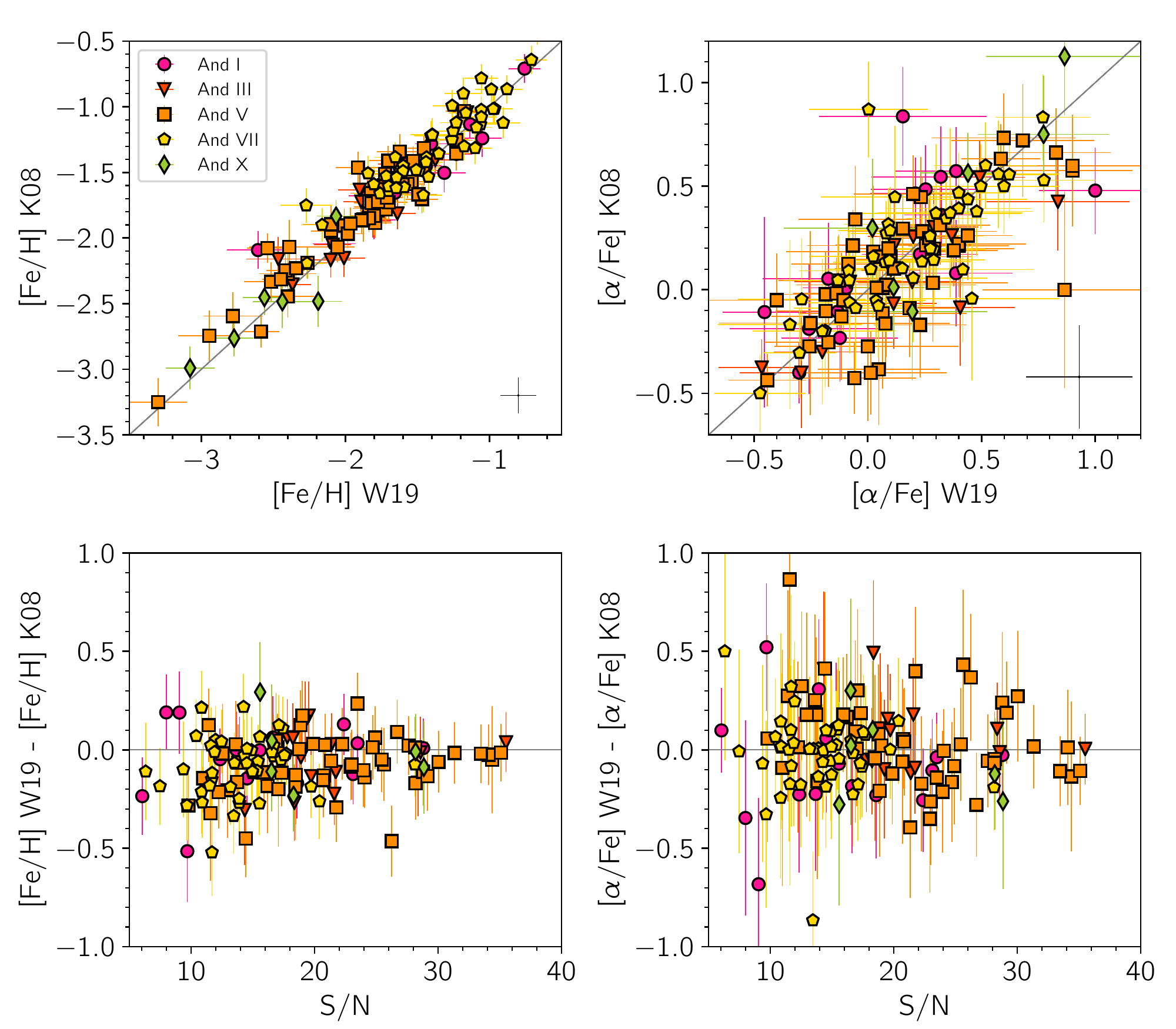}
    \caption{The same as Figure~\ref{fig:ie_vs_enk_gcs}, but for our sample of dSphs with deep observations.}
    \label{fig:ie_vs_enk_dsphs}
\end{figure*}

Because we use a revised version \citep{escala18}, rewritten in Python, of the pipeline outlined in \citet{kirby08a}, we verify that our results are consistent with those obtained using the IDL version from \citet{kirby08a}. To do this, we recompute \feh\ and \alphafe\ for our GC and deep dSph samples using the most up-to-date version of the IDL pipeline. Figures~\ref{fig:ie_vs_enk_gcs} and \ref{fig:ie_vs_enk_dsphs} show the comparison between [Fe/H] and [$\alpha$/Fe] from both pipelines, and their associated total uncertainties for our sample of MW GCs and M31 dSphs with deep observations. For our GC sample ($N_{\mathrm{stars}} = 196$, Figure~\ref{fig:ie_vs_enk_gcs}), we find a median difference of $-0.06\pm0.16$ in \feh, and a median difference of $-0.01\pm0.12$ in \alphafe. The standard deviation for all GCs is $0.08\pm0.01$ and $0.10\pm0.02$ for \feh\ and \alphafe, respectively. To quantify the agreement between the two pipelines, we follow the formalism described in the Appendix of \citet{kirby20}. We compute the standard deviation of the difference between the pipelines, normalized by their uncertainties, as follows: 
\begin{equation}
    \label{eq:stddev_weighted}
    \mathrm{stddev}\left(\frac{X_{W19} - X_{K08}}{\sqrt{\delta X_{W19}^2 + \delta X_{K08}^2}}\right) 
\end{equation}

\noindent where $X$ is either \feh\ or \alphafe, and $\delta X$ is the corresponding total uncertainty. For our GCs, we find this value to be $0.51\pm0.07$ for \feh, and $0.55\pm0.07$ for \alphafe. A value of 1.0 or less would indicate that the uncertainties completely explain the scatter in the difference, where a value larger than 1.0 indicates an additional source of uncertainty. We conclude that, for GCs, the pipelines agree within their respective uncertainties because the standard deviations from Equation~\ref{eq:stddev_weighted} are less than 1.0.

For our M31 dSphs with deep spectroscopic observations ($N_{\mathrm{stars}} = 145$, Figure~\ref{fig:ie_vs_enk_dsphs}), we find a median difference of $-0.06\pm0.12$ for \feh, and $0.00\pm0.22$ for \alphafe, with a standard deviation of $0.15\pm0.02$ for \feh\ and $0.22\pm0.04$ for \alphafe. Applying Eq.~\ref{eq:stddev_weighted} to our sample of dSphs, we find values of $0.78\pm0.09$ for \feh\ and $0.60\pm0.09$ for \alphafe. As in the case of the GCs, these values are less than 1.0, and we confirm that the pipelines agree for individual \feh\ and \alphafe\ measurements from our dSphs sample.

\section{Validation coadds grouped by photometric effective temperature}
\label{sec:bin_teff}
In addition to grouping our stars according to their photometric metallicity ([Fe/H]$_{\rm phot}$), we also grouped our stars according to their photometric effective temperature. We prefer to use the [Fe/H]$_{\rm phot}$ grouping for our analysis because the spectral synthesis technique that we use can more easily account for a wide range of \teff\ than \feh\ \citep{yang13}. By design, we use the values of \teff\ measured from photometry in constructing the synthetic spectrum (Section~\ref{sec:measuring_abundances_coadd} that our coadded spectrum is compared to. Therefore, our method accounts for a potentially wide range of \teff\ for a given coadd when binned by \feh\. The same cannot be said for the \teff\ binning scenario, where stars with a large range of \feh$_{\mathrm{phot}}$ may be coadded together, as we fit for \feh.

\begin{figure*}
	\includegraphics[width=0.95\textwidth]{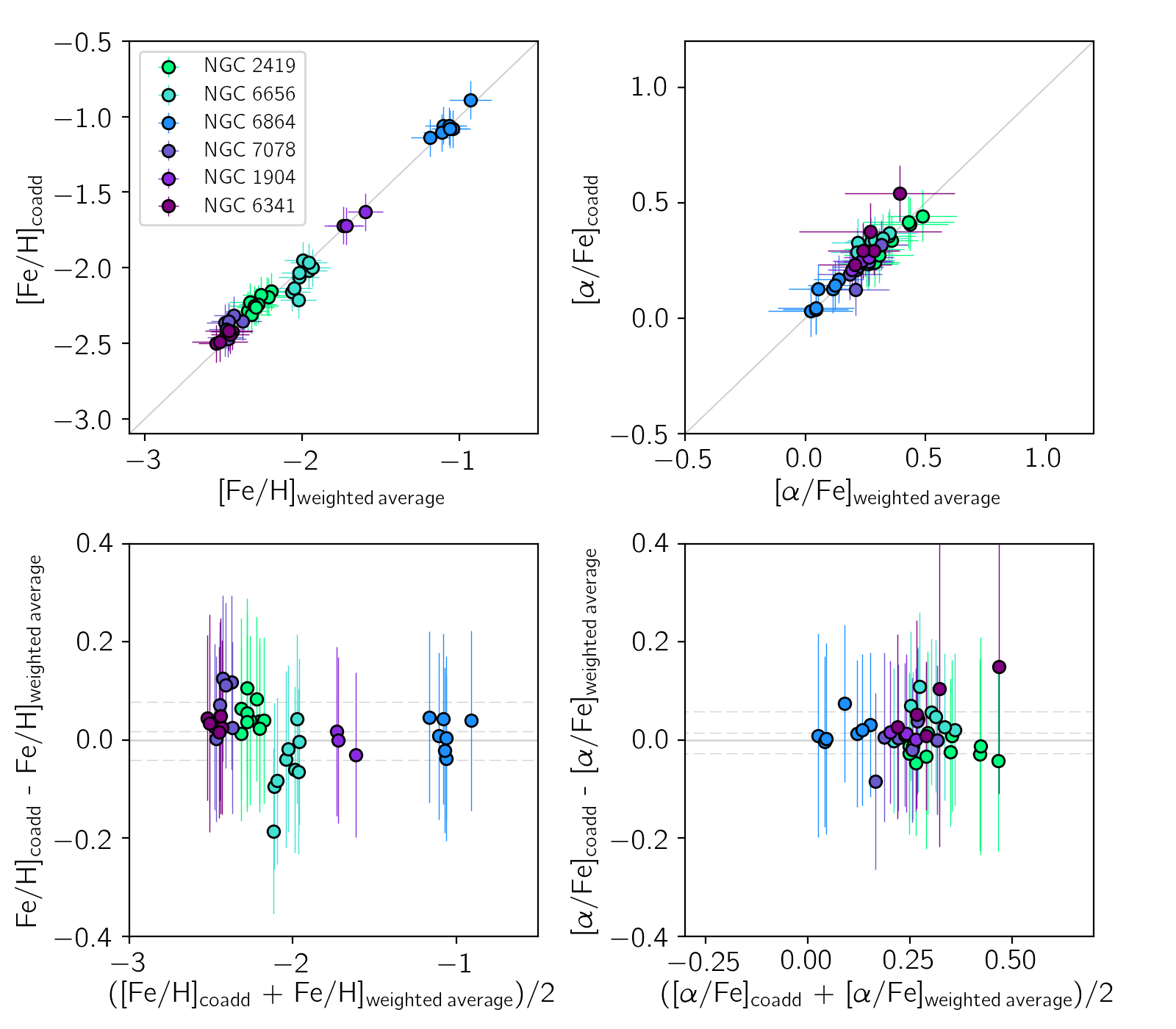}
    \caption{The same as Figure~\ref{fig:weighted_average_comparison_gcs}, but for stars binned by \teff.}
    \label{fig:weighted_average_comparison_gcs_bin_teff}
\end{figure*}

\begin{figure*}
	\includegraphics[width=0.95\textwidth]{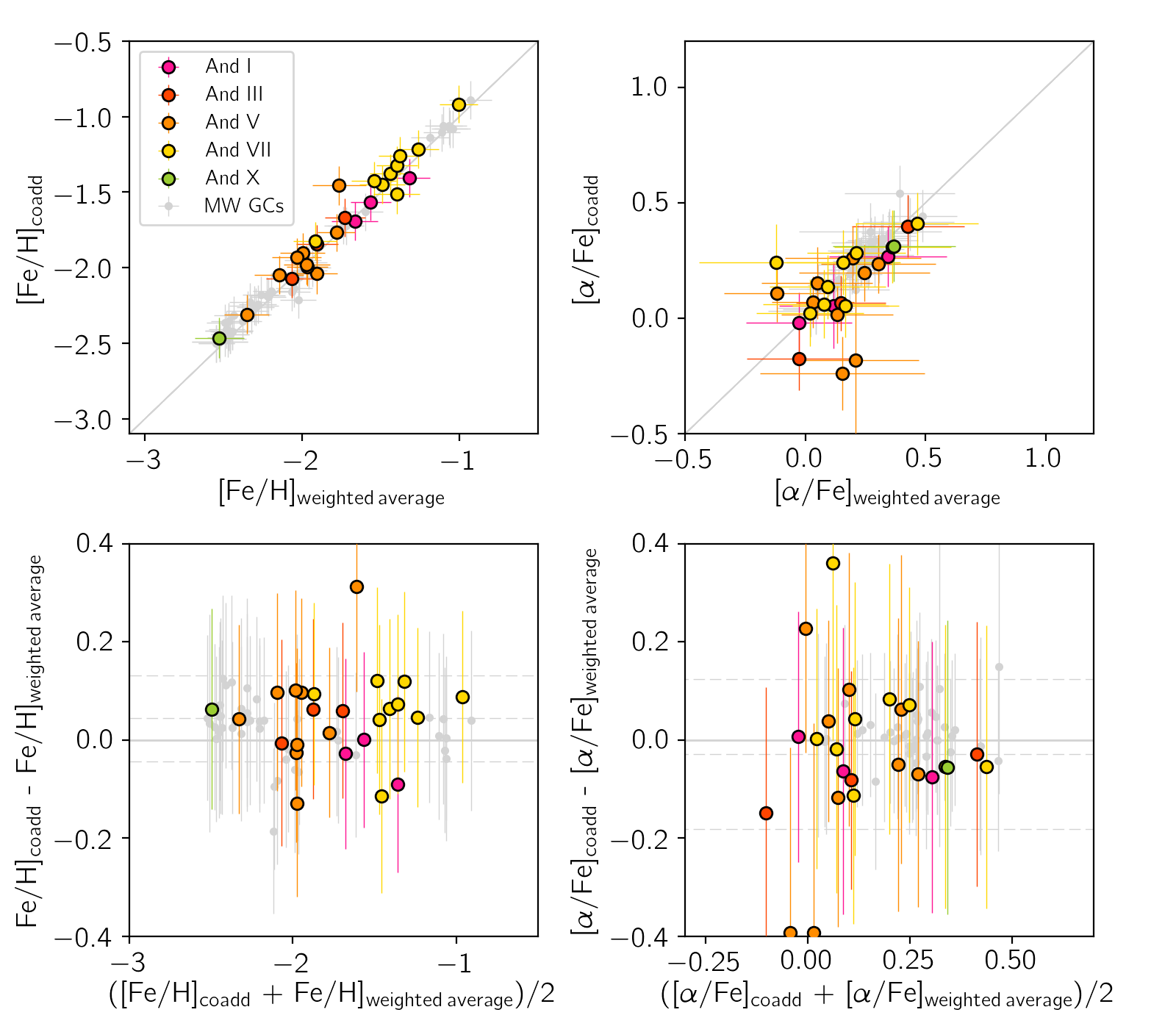}
    \caption{The same as Figure~\ref{fig:weighted_average_comparison_deep_dsphs}, but for stars binned by \teff.}
    \label{fig:weighted_average_comparison_deep_dsphs_bin_teff}
\end{figure*}

\begin{figure*}
	\includegraphics[width=0.95\textwidth]{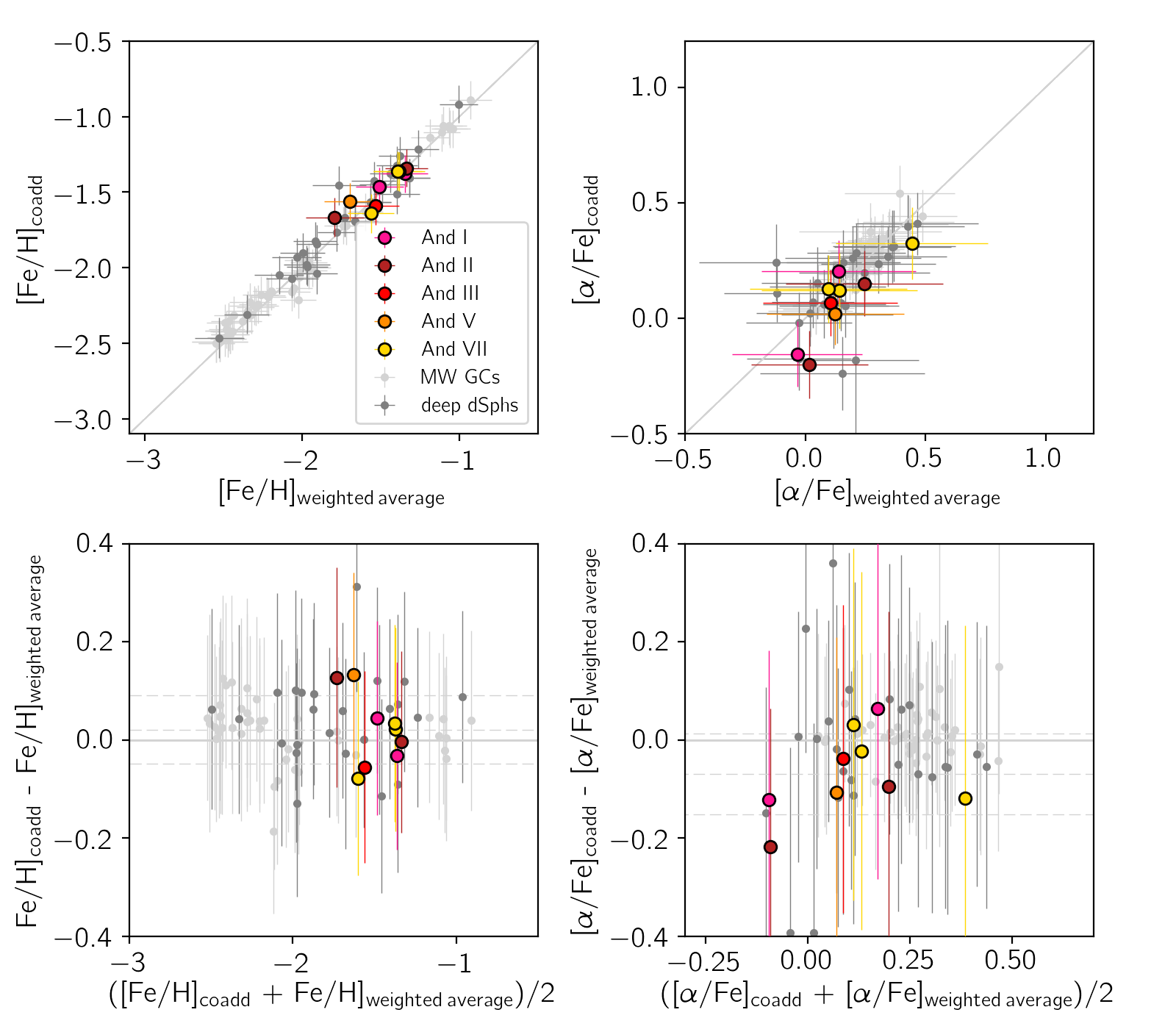}
    \caption{The same as Figure~\ref{fig:weighted_average_comparison_shallow_dsphs}, but for stars binned by \teff.}
    \label{fig:weighted_average_comparison_shallow_dsphs_bin_teff}
\end{figure*}

In Figures~\ref{fig:weighted_average_comparison_gcs_bin_teff}, \ref{fig:weighted_average_comparison_deep_dsphs_bin_teff}, and \ref{fig:weighted_average_comparison_shallow_dsphs_bin_teff}, we find qualitatively similar results to those shown in the main text. We find no significant difference between the binning scenarios in either the values of the offsets or dispersion for all three of our samples. 

\begin{figure*}
\centering
	\includegraphics[width=0.8\textwidth]{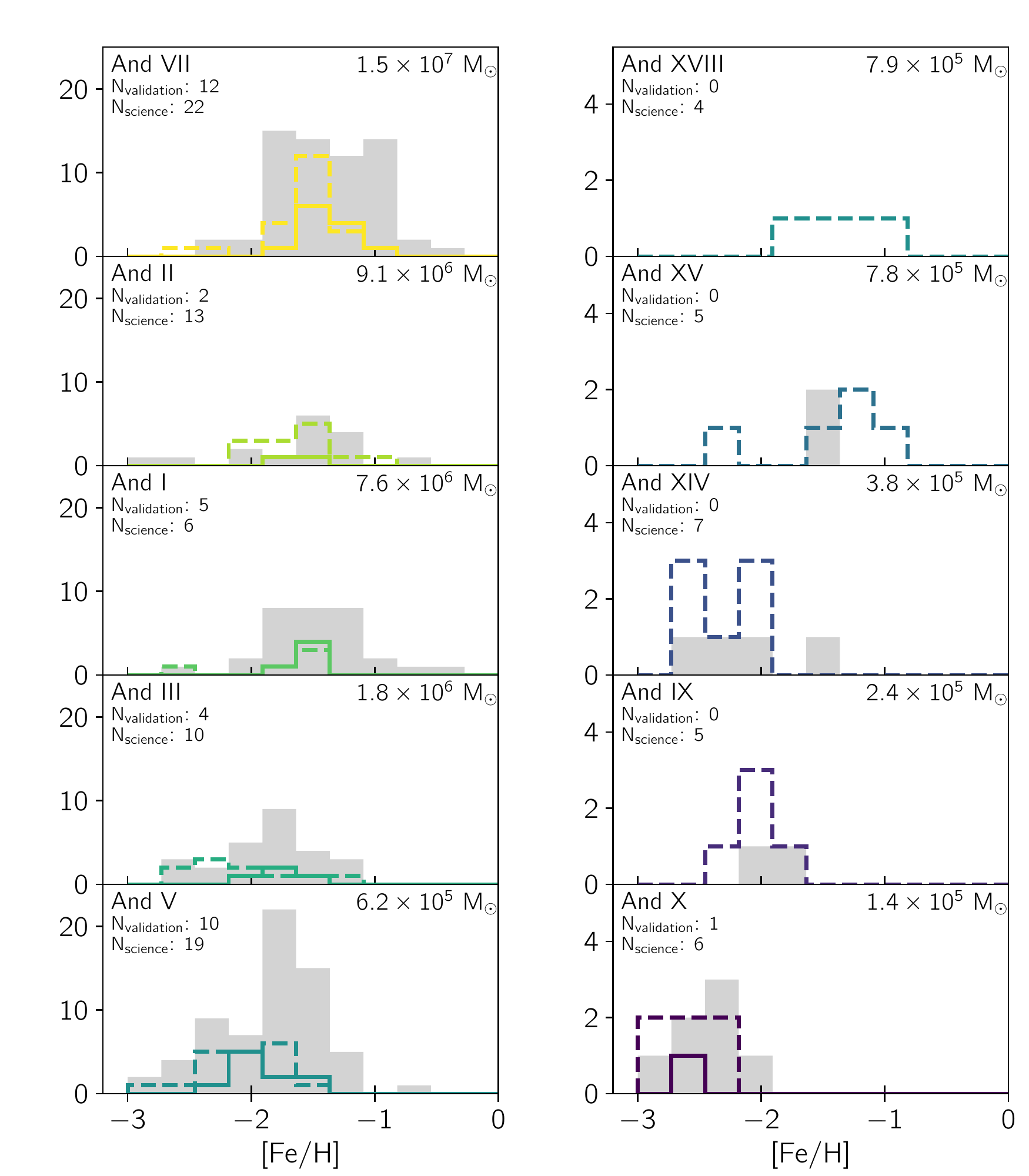}
    \caption{The same as Figure~\ref{fig:good_vs_failed_feh_hist}, but for stars binned by \teff.
    }
    \label{fig:good_vs_failed_feh_hist_teff}
\end{figure*}

\begin{figure*}
\centering
	\includegraphics[width=0.8\textwidth]{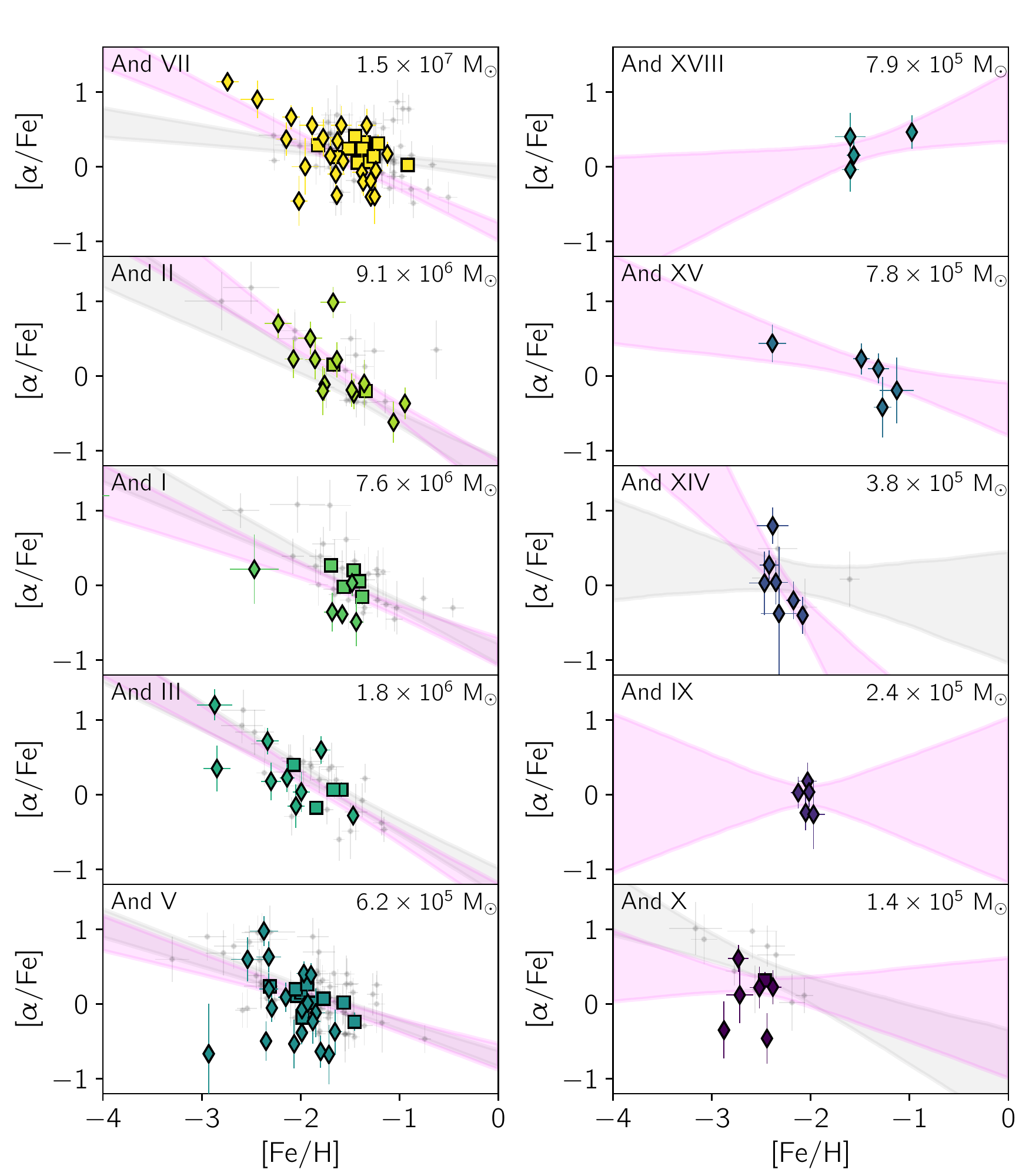}
    \caption{The same as Figure~\ref{fig:feh_alphafe_dsphs}, but for stars binned by \teff.} 
    \label{fig:feh_alphafe_dsphs_teff}
\end{figure*}

In Figure~\ref{fig:good_vs_failed_feh_hist_teff} we show the \feh\ distribution for our sample of individual stars, validation coadds, and science coadds, for each of our ten M31 dSphs. We find these distributions to be very similar to those shown in Figure~\ref{fig:good_vs_failed_feh_hist}, using the [Fe/H]$_{\rm phot}$ binning scenario. In Figure~\ref{fig:feh_alphafe_dsphs_teff} we show the 2D \alphafe-\feh\ distribution for our M31 dSphs, again finding similar results as with the coadds binned by [Fe/H]$_{\rm phot}$. {One notable exception is And VII, where there appears to be a stronger correlation between \feh\ and \alphafe\ than seen in Figure~\ref{fig:feh_alphafe_dsphs}. However, we refrain from drawing any further conclusions from the differences in And VII, as the  slope in Figure~\ref{fig:feh_alphafe_dsphs_teff} is predominately due to only two points (both coadds).} As expected, binning by $T_{\rm eff,phot}$ results in a narrower range of [Fe/H] values compared to binning by ${\rm [Fe/H]}_{\rm phot}$. We draw the same conclusion as \citet{yang13}: both binning scenarios produce abundance distributions that agree within their associated uncertainties. Our decision to use the [Fe/H]$_{\rm phot}$ is motivated primarily by the desire to limit potential sources of additional systematic uncertainties, as a range of \teff\ in a given [Fe/H]$_{\rm phot}$ bin is more easily accounted for in our pipeline.

\end{document}